%% file: main.tex
\documentclass[a4paper,fleqn]{cas-sc}

\usepackage{hyperref}
\usepackage{bm}
\usepackage{mathtools}
\usepackage{subcaption}
\usepackage{tikz}
\usepackage{pgfplots}

\usepackage[numbers]{natbib}

\graphicspath{{figures/}}

\newcommand\wij[3][w]{#1_{#2,#3}}

\newcommand\der[3]{d^{(#1)}_{#2;\,i,j}(#3)}

\newcommand\lin{\mathrm{lin}}
\newcommand\dd{\mathrm{d}}
\newcommand\rme{\mathrm{e}}
\begin{document}

\shorttitle{A simple and efficient second-order immersed-boundary method}
\shortauthors{}
\title [mode = title]{A simple and efficient second-order immersed-boundary method for the incompressible Navier--Stokes equations}

\author[1]{Paolo Luchini}[orcid=0000-0001-6527-7762]
\ead{luchini@cplcode.net}
\credit{Conceptualization, Methodology, Formal analysis, Software, Writing - Original Draft, Writing - Review \& Editing, Supervision}

\author[2]{Davide Gatti}[orcid=0000-0002-8178-9626]
\ead{davide.gatti@kit.edu}
\credit{Methodology, Investigation, Validation, Formal analysis, Writing - Original Draft}

\author[3,4]{Alessandro Chiarini}[orcid=0000-0001-7746-2850]
\ead{alessandro.chiarini@polimi.it}
\credit{Methodology, Investigation, Validation, Formal analysis, Writing - Original Draft}

\author[4]{Federica Gattere}[orcid=0000-0002-1450-415X]
\ead{federica.gattere@polimi.it}
\credit{Methodology, Investigation, Validation, Formal analysis, Writing - Original Draft}

\author[4]{Marco Atzori}[orcid=0000-0003-0790-8460]
\ead{marco.atzori@polimi.it}
\credit{Methodology, Investigation, Validation, Formal analysis, Writing - Original Draft}

\author[4]{Maurizio Quadrio}[orcid=0000-0002-7662-3576]
\ead{maurizio.quadrio@polimi.it}
\credit{Methodology, Investigation, Formal analysis, Writing - Original Draft, Writing - Review \& Editing, Supervision, Project administration}
\cormark[1]

\affiliation[1]{
                organization={Dip. di Ingegneria Industriale, Universit\'a di Salerno},
                addressline={via Giovanni Paolo II 132},
                city={Fisciano},
                postcode={84084},
                country={Italy}
}
\affiliation[2]{
                organization={Institute of Fluid Mechanics, Karlsruhe Institute of Technology},
                addressline={Kaiserstra\ss e 10},
                city={Karlsruhe},
                postcode={76131},
                country={Germany}
}
\affiliation[3]{
                organization={Complex Fluids and Flows Unit, Okinawa Institute of Science and Technology Graduate University},
                addressline={1919-1 Tancha, Onna-son},
                city={Okinawa},
                postcode={904-0495},
                country={Japan}
}
\affiliation[4]{
                organization={Dip. di Scienze e Tecnologie Aerospaziali, Politecnico di Milano},
                addressline={Campus Bovisa, via La Masa 34},
                city={Milano},
                postcode={20156},
                country={Italy}
}

\cortext[1]{Corresponding author}


\begin{abstract}
An immersed-boundary method for the incompressible Navier--Stokes equations is presented. It employs discrete forcing for a sharp discrimination of the solid-fluid interface, and achieves second-order accuracy, demonstrated in examples with highly complex three-dimensional geometries.
The method is implicit, meaning that the point in the solid which is nearest to the interface is accounted for implicitly, which benefits stability and convergence properties; the correction is also implicit in time (without requiring a matrix inversion), although the temporal integration scheme is fully explicit.
The method stands out for its simplicity and efficiency: when implemented alongside second-order finite differences, only the weight of the center point of the Laplacian stencil in the momentum equation is modified, and no corrections for the continuity equation and the pressure are required. 
The immersed-boundary method, its performance and its accuracy are first verified on simple problems, and then put to test on a simple laminar, two-dimensional flow and on two more complex examples: the turbulent flow in a channel with a sinusoidal wall, and the flow in a human nasal cavity, whose extreme anatomical complexity mandates an accurate treatment of the boundary.
\end{abstract}



\maketitle

\input{introduction}
\input{method}

\input{discussion}
\input{results}
\input{conclusions}

\section*{Declaration of competing interest}
The authors declare that there are no conflicts of interest.

\section*{Acknowledgments}
This research has been partially supported by ICSC -- Centro Nazionale di Ricerca in High Performance Computing, Big Data, and Quantum Computing funded by European Union --NextGenerationEU. European Union support through Next Generation EU, Mission 4, Component 1, CUP D53D2300343006 // PRIN 2022 - OpenNOSE is gratefully acknowledged. Part of the computer time has been provided by CINECA through the ISCRA B project OptiNose.


\printcredits

\bibliographystyle{cas-model2-names}
\bibliography{../Wallturb,../Nose}

\end{document}

%% file: introduction.tex
\section{Introduction}
\label{sec:introduction}

Immersed-boundary methods (IBMs) have seen their popularity increase over the last two decades, and are nowadays often employed in the numerical simulation of fluid flows around complex geometries.
They represent an interesting alternative to the classic methods which discretize the fluid equations on a body-conforming grid, and are particularly well suited to situations where the solid bodies have a complex shape, move or deform. 
An IBM relies on a Cartesian grid, where grid points generally do not coincide with the contours of the bodies. 
Using a Cartesian grid brings along substantial advantages compared to body-conforming grids: easier generation of a structured mesh, simpler and more efficient solution algorithms and parallelisation, savings in memory requirements and computing time. 
IBMs may render problems affordable in complex and/or moving geometries which would otherwise be prohibitively expensive from the point of view of the computational complexity, typical examples being those involving fluid-structure interactions and/or bio-medical applications \cite{detullio-pascazio-2016, griffith-patankar-2020}, or particles-laden flows \cite{uhlmann-2005, zhu-etal-2024}.
The obvious drawback is that the boundary conditions on the body are defined at locations that in general do not coincide with grid points; they are therefore enforced at grid points, by either altering the volume forces or interpolating velocity values near the boundary, which can be thought of as being ``immersed'' in the fluid. 

IBMs can be traced back to the seminal work of Peskin \cite{peskin-1972}, and were extended over the years, in particular starting from Ref. \cite{fadlun-etal-2000}. Comprehensive reviews are provided in Refs. \cite{peskin-1972, iaccarino-verzicco-2003, mittal-iaccarino-2005, sotiropoulos-yang-2014} and in the recent contributions \cite{verzicco-2023, mittal-seo-2023}.
IBMs are generally categorised into two classes \cite{mittal-iaccarino-2005}, depending on whether they are based on a continuous or discrete forcing. 
The continuous-forcing IBM adds a volume forcing term to the continuous Navier--Stokes equations before discretization. 
Examples of this class of IBM are described in Refs. \cite{peskin-1972, goldstein-handler-sirovich-1993, saiki-biringen-1996}. 
Such IBMs have been used in different biological and engineering applications \cite{fauci-peskin-1988, zhu-peskin-2002, kim-peskin-2007, kim-lai-2010}, and various forcing functions have been proposed. 
The continuous-forcing IBM, however, unavoidably suffers from the actual boundary being smeared over several nearby grid points because of the forcing function, and from the need to derive {\em ad hoc} forcing parameters. 
Moreover, the governing equations need to be solved in the whole domain, including within the solid body, which leads to an aggravation of their computational cost. 
The second class of methods, referred to as the discrete-forcing IBM, applies the forcing (either explicitly or implicitly) to the already discretised Navier--Stokes equations; examples can be found in Refs. \cite{ye-etal-1999, fadlun-etal-2000, balaras-2004, orlandi-leonardi-2006, chi-lee-im-2017}. 
A sharp representation of the boundary, leading to the same accuracy as a body-conforming grid \cite{mittal-seo-2023}, becomes possible with the discrete-forcing IBM; however, since the forcing is only introduced after discretization, such IBMs are tightly linked to the underlying spatial and temporal discretization of the flow solver. 

The present work describes a discrete-forcing IBM. In early attempts, e.g. Ref. \cite{fadlun-etal-2000}, the discrete forcing in the momentum equation was computed on the body surface and inside the body as well, while an additional explicit source term needed to restore mass conservation near the boundaries was computed in a later step. 
An alternative approach, called ghost-node IBM, was introduced in \cite{fedkiw-2002} and further extended over the years \cite{tseng-ferziger-2003, ghias-mittal-dong-2007, mittal-etal-2008, li-etal-2023}. 
Ghost nodes are those grid nodes that lie in the solid but at the same time belong to the stencil used to discretize differential operators appearing in the governing equations at fluid points. 
With ghost nodes, the forcing can be introduced implicitly in the momentum equations by means of the discrete stencil operators. Therefore, the number of ghost node layers depends on the discretisation. 
The general idea of a ghost-node IBM is to enforce the boundary conditions by means of the values of the variables at the ghost nodes; these are extrapolated from values at the internal points and from those at the boundary, known from the boundary conditions.
Typically, a single value of each flow variable is associated to each ghost node, and is extrapolated along the direction normal to the boundary. 
Over the years several extrapolation schemes have been proposed. 
Mittal et al. \cite{mittal-etal-2008} and Ghias et al. \cite{ghias-mittal-dong-2007} used linear extrapolation to obtain a second-order convergence. 
The same convergence was obtained by Tseng at al. \cite{tseng-ferziger-2003}, who employed a quadratic extrapolation, and by Gao et al. \cite{gao-tseng-lu-2007} via a second-order Taylor series expansion. 
Employing a wider stencil near the boundary leads to a higher order of convergence (see e.g. Ref. \cite{seo-mittal-2011}).
Recently, however, Chi et al. \cite{chi-etal-2020} observed that, regardless of the reconstruction method, boundary conditions in the various directions cannot be accurately and simultaneously represented by a single ghost-node value; in their IBM they define and compute multiple ghost-node values, one for each direction.

In a conventional ghost-node IBM, multiple grid points in the fluid are used to extrapolate the ghost-node value \cite{tseng-ferziger-2003, gao-tseng-lu-2007}. 
In Refs. \cite{pan-shen-2009, chi-lee-im-2017, chi-etal-2020}, a single fluid point is considered; however, this is chosen as the second fluid point instead of the closest to the boundary, in order to avoid numerical instability issues arising when the distance between the first point and the boundary tends to zero. 
This is because most existing implementations deal with the forcing term explicitly, and require the computation of the ghost-node values at each iteration, thereby increasing the overall computational cost. The equivalent implicit treatment is typically not pursued, because it involves a matrix inversion. 
A further drawback brought about by a wide interpolation stencil arises when the thickness of the body is locally less than the size of the local grid spacing, or when two surfaces are separated by a number of points which is less than the stencil width.
Interpolating over a wider stencil also entails delivering a worse approximation of the solution near the boundary.

This paper introduces a simple and computationally efficient, implicit in space and time, second-order accurate IBM for the incompressible Navier--Stokes equations, based upon and tightly integrated with a second-order finite difference method. Differently from several discrete-forcing IBMs, a boundary condition for pressure is not required. The time integration scheme is generic but of the explicit type, a design choice motivated by our interest towards large-scale simulations of turbulent flows.

In the steady case, the implicit treatment of the ghost nodes is computationally cheap and sufficient to restore the diagonal dominance of the whole system, which helps in the iterative solution process. 
It consists in a modification of the weight of the midpoint of the Laplacian stencil, under the assumption that close to the boundary the viscous terms are dominant. 
Besides avoiding to compute and store the solution at the ghost node, as typical of other IBMs, the formulation is free from the numerical instabilities arising when the distance between the first point in the fluid and the boundary vanishes. 
Once the problem becomes unsteady, the immersed-boundary correction, when treated implicitly in time, still preserves the same diagonal dominance in the iterative solution, and does not affect the stability limit of the explicit time integration scheme. 

The formulation of our IBM originally combines various aspects, some of which are of course already available in previous work. For example, extrapolating the ghost-node values via a linear formula that includes the boundary point and the first fluid point was already done, among others, by Gibou et al. in Ref.\cite{gibou-etal-2002}, in the context of the variable-coefficients Poisson equation. We also share with Ref.\cite{gibou-etal-2002} the use of different ghost-node values for each Cartesian direction; multiple ghost nodes were used also in Ref.\cite{chi-etal-2020}, although with an explicit treatment.
Modifying only the Laplacian close to the boundary, as well as computing distinct coefficient values for each direction, was also suggested by Orlandi \& Leonardi in Ref. \cite{orlandi-leonardi-2006}; however, their correction was not limited to the central point of the stencil, which is a requisite for the corrections in different directions to be additive, and the velocity at the ghost nodes was set to zero.

The paper is organised as follows. After this Introduction, Sec. \ref{sec:IBM} thoroughly describes our IBM, its general design and then its implementation into a Navier--Stokes finite-difference solver, considering first the steady and then the unsteady case. A critical discussion of the IBM and of its advantages and drawbacks is offered in Sec. \ref{sec:discussion}, where Neumann-type boundary conditions are mentioned. Finally, Sec. \ref{sec:results} provides an exhaustive discussion of accuracy and performance of the method as applied to three example flows where geometrical complexity is significant. The paper concludes with some final remarks drawn in Sec. \ref{sec:conclusions}.

%% file: method.tex
\section{The immersed-boundary method}
\label{sec:IBM}
The IBM is introduced in the context of elementary, linear flow problems: a steady flow example is used first to illustrate the spatial accuracy of the method, followed by an example that involves a time-dependent flow, where the temporal accuracy of the method is discussed.

The starting point is a basic ``staircase'' approximation of the boundary, in which each point is defined as ``internal'' (in the fluid region) or ``external'' (in the solid region); this first-order representation is improved upon via the IBM correction, which gets applied only to those stencils of the discretised equations which cross the solid boundary. In fact, the method can be alternatively interpreted as a deferred correction, a sometimes helpful viewpoint.
The correction exploits the fact that, near the boundary, viscous terms dominate over convective and pressure terms. Therefore, the presence of the boundary is accounted for simply by altering the weight of the central point of the star-shaped stencil of the 7-point Laplacian at the first internal point: no additional corrections of the convective terms are required (albeit they are possible) to achieve the second order. In addition, as in some other implementations, no correction for the pressure or the continuity equation needs to be expressly introduced; the absence of such need can be explained by the lack of a boundary condition for pressure in the continuous Navier--Stokes problem: when there is no boundary condition there is no position where to impose it. While IBMs without an express pressure boundary correction have been used before, to the best of our knowledge this is the first time that the validity of such choice is {\it a posteriori} confirmed through convergence tests (see Sec. \ref{sec:results}).

\subsection{Equations of motion and discretization}
\label{sec:discretization}

The IBM is implemented in a solver for the direct numerical simulation (DNS) of the incompressible Navier--Stokes equations, written in primitive variables and with suitable initial and boundary conditions:

\begin{equation}
\begin{dcases} 
\frac{\partial \bm{u}}{\partial t} + \left( \bm{u} \cdot \bm{\nabla} \right) \bm{u} - \frac{1}{Re} \nabla^2 \bm{u} + \bm{\nabla} p = \bm{g} \\
\bm{\nabla} \cdot \bm{u}=0 
\end{dcases}
\label{eq:ns}
\end{equation}
where $t$ is the time, $\bm{u}$ is the velocity vector, $p$ is the reduced pressure, and $\bm{g}$ is a possible body force. $Re=U_{ref}L_{ref}/\nu$ is a Reynolds number built with the reference velocity $U_{ref}$, the reference length $L_{ref}$ and the kinematic viscosity $\nu$ of the fluid. In a Cartesian frame, the spatial directions are denoted as $x,y$ and $z$, and the corresponding velocity components as $u,v$ and $w$.

\begin{figure}
\centering
\includegraphics[scale=0.45]{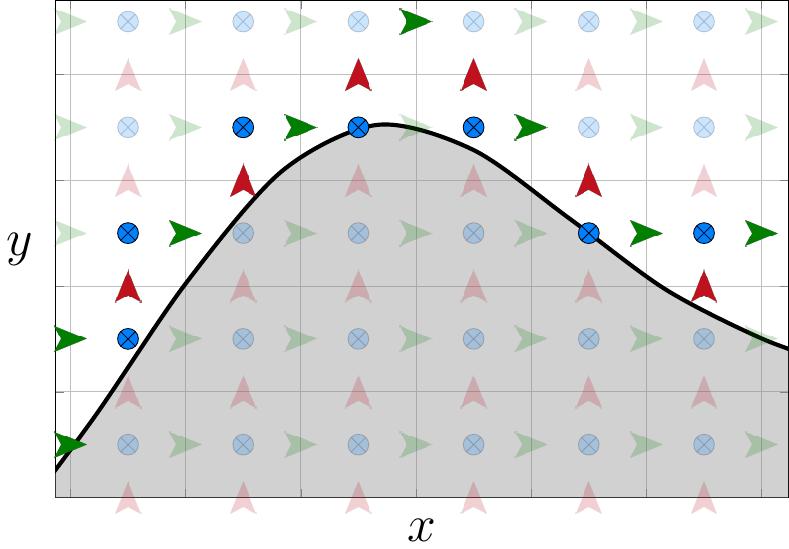}
\caption{Two-dimensional view of a solid body (gray background) immersed in a fluid (white background), with an overlaid staggered Cartesian grid. The collocation points for the velocity components in the $x$, $y$ and $z$ directions are drawn in green, red and blue. Dull colors denote the fully internal/external points, and vivid colors the points where the immersed-boundary correction is applied.}
\label{fig:staggered-grid}
\end{figure}

The Navier--Stokes equations are advanced in time using a standard incremental pressure-correction scheme coupled with a fully explicit temporal integration scheme. 
The momentum equation is first advanced in time without the incompressibility constraint, that is enforced later during the so-called projection step. The velocity field gets projected onto a solenoidal vector field and the required pressure increment is found by (exactly or approximately) solving a Poisson equation, and then used to update the pressure field.

The spatial discretization takes place on a Cartesian grid that is staggered in the three directions, as sketched in figure \ref{fig:staggered-grid}. Pressure is defined at the center of each cell, whereas each velocity component is defined at the relative interface. Uniform as well as non-uniform spacing is possible in each direction.

The discretisation of the differential operators relies on centered second-order central finite differences in every direction, with a three-points stencil for each direction. To introduce the notation used in the rest of the paper, let us consider a two-dimensional case for simplicity. A generic grid point of coordinates $(x,y)$ is identified by the pair of integers $(i,j)$, such that $x=x_i$ and $y=y_j$. Taking the first derivative as an example, for a scalar function $f$ the first derivative along the $x$ direction at the $(i,j)$ position is written as:
\begin{equation}
\left. \frac{\partial f}{\partial x} \right|_{i,j} = 
  \der{1}{x}{-1} f_{i-1,j} + \der{1}{x}{0} f_{i,j} + \der{1}{x}{1} f_{i+1,j}
  + \mathcal{O}(\Delta x^2) \ .
\end{equation}
In the expression above, the three symbols $\der{1}{x}{\cdot}$ indicate the three finite-differences coefficients for the centered first derivative along the $x$ direction, evaluated at point $(i,j)$.

\subsection{The steady case}
\label{sec:steady}

\begin{figure}
\centering
\includegraphics[]{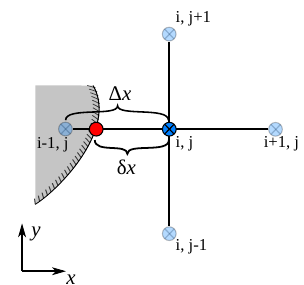}
\caption{Computational stencil for the $w$ velocity component in the $x-y$ plane. As in figure \ref{fig:staggered-grid}, vivid blue highlights the point where the immersed-boundary correction is applied. A red dot denotes the actual boundary intersection.}
\label{fig:stencil}
\end{figure}

We start by considering the steady case, and in particular the Laplacian operator: since our IBM relies on the Laplacian being the dominant term near the solid boundary, this simple example is particularly significant.
At a generic $(i,j)$ position the Laplacian of, e.g., the $z$ velocity component  discretised by second-order central finite differences reads:
\begin{equation}
\nabla^2 w \approx
    \der{2}{x}{-1} \wij{i-1}{j} + \der{2}{x}{0} \wij{i}{j} + \der{2}{x}{1} \wij{i+1}{j} +
    \der{2}{y}{-1} \wij{i}{j-1} + \der{2}{y}{0} \wij{i}{j} + \der{2}{y}{1} \wij{i}{j+1} .
\label{eq:disclapl}
\end{equation}

Let us assume, as shown in figure \ref{fig:stencil}, that point $(i,j)$ is close to the solid body; the boundary crosses the left arm of the computational stencil, and the left neighbour of point $(i,j)$, i.e. point $(i-1,j)$, lies within the solid. 
The simplest description of the boundary is achieved by setting the velocity at the external (in the solid region) point to zero, i.e. $\wij{i-1}{j}=0$. This amounts to a staircase, i.e. piecewise parallel to the axes, approximation of the boundary, whose maximum error in the position of the body contour is proportional to $\Delta x = x_{i,j} - x_{i-1,j}$, i.e. one of first-order accuracy.

To increase the accuracy of the description and to avoid deteriorating the overall second-order accuracy of the underlying numerical method, the representation of the boundary needs to be improved to a piecewise-linear approximation. 
The velocity at a generic point $x$ can be computed as a linear interpolation between the velocity $\wij{i}{j}$ at the first fluid point and $w=0$ at the true boundary, located at $x_i - \delta x$.
The linear function that fits velocity between the position $(i,j)$ and the true boundary reads:
\begin{equation}
  \lin(x)= \left( 1 + \frac{ x - x_i }{\delta x} \right) \wij{i}{j} \, 
  \label{eq:interp-lin}
\end{equation}
which satisfies $\lin(x_i)=\wij{i}{j}$ and $\lin(x_i-\delta x)=0$.

The same function can be used to linearly extrapolate (instead of setting it to zero) the value $\wij{i-1}{j}$, corresponding to the ghost point that falls inside the solid region and that is needed for building the stencil of the discretised Laplacian \eqref{eq:disclapl} in the $(i,j)$ position. The extrapolated value reads:
\begin{equation}
  \lin(x_{i-1})= \left( \frac{ \delta x - \Delta x }{\delta x} \right) \wij{i}{j} \, .
  \label{eq:interp-lin-extr}
\end{equation}

The above is similar to other implementations of the IBM. Crucial to ours is the observation that the extrapolated value for $\wij{i-1}{j}$ does not need to be stored explicitly as a ghost value, but can be substituted back into equation \eqref{eq:disclapl} and accounted for implicitly (the same will remain true in unsteady problems). 
The substitution concentrates the modification in the coefficient of the central point of the stencil, leading to a single value to be stored (this remains the case even when the correction needs to be applied along multiple directions). The updated coefficient $\tilde{d}^{(2)}_{x;i,j}(0)$ reads:
\begin{equation}
\tilde{d}^{(2)}_{x;ij}(0) \equiv
    \left( \der{2}{x}{0} - \der{2}{x}{-1} \frac{\Delta x -  \delta x}{\delta x} \right)
\label{eq:disclapl-extr}
\end{equation}
where the term 
\begin{equation}
\der{2}{x}{-1} \frac{\Delta x -  \delta x}{\delta x}
\label{eq:lambda}
\end{equation}
embodies the immersed-boundary correction, referred to in the following as $\lambda$. It is worth noting that, owing to the opposite signs of $\der{2}{x}{0}$ and $\der{2}{x}{-1}$, the updated coefficient $\tilde{d}^{(2)}_{x;i,j}(0)$ is always of the same sign; its absolute value monotonically increases for $\Delta x>\delta x>0$, and can not be zero. This leads to an increased diagonal dominance, more and more so when $\delta x \rightarrow 0$. 

The extension to the general case, in which the immersed boundary crosses more than one arm of the computational stencil (i.e. more than one neighbouring point is within the solid), is straightforward. The corrections to the central point of the Laplacian ensuing from derivatives in different directions are additive, and can be computed independently. 


\subsection{The steady case: example}
\label{sec:example-steady}

The potential of the IBM is illustrated through a simple example. We consider the laminar parallel flow in a circular pipe in the absence of external volume forces. If the pipe axis (or, generally, the axis of any straight duct with arbitrarily shaped cross section) is aligned with the $z$ direction of the Cartesian reference system, the problem is homogeneous along $z$, and reduces to a two-dimensional and steady problem in the $(x,y)$ plane, where the only differential operator is a Laplacian. 
The governing equations in fact simplify to the following Poisson equation:
\begin{equation}
\frac{1}{Re_b} \nabla^2 w = \frac{\partial p}{\partial z} \, ,
\label{eq:lapl}
\end{equation}
where $w$ is the streamwise velocity component, which does not depend on the streamwise coordinate, $Re_b$ is the Reynolds number, and on the right-hand side the spatially uniform pressure gradient has the value $\partial p / \partial z = \Pi$.

\begin{figure}
\centering
\includegraphics[]{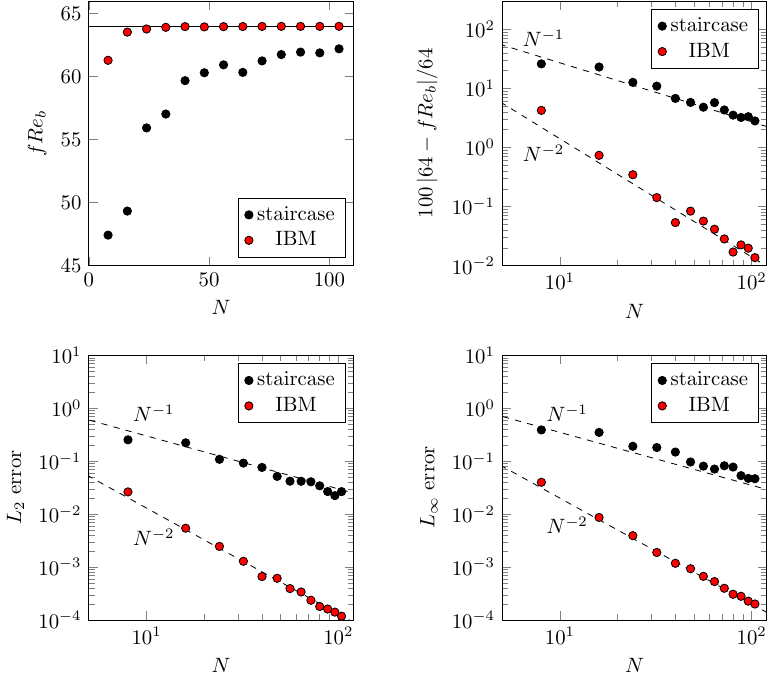}
\caption{Laminar Hagen--Poiseuille flow, equation \eqref{eq:lapl}. Top left: value of $f Re_b$ as computed on a $N \times N$ Cartesian grid (exact value $f Re_b = 64$ is the horizontal line). Top right: percentage error of $f Re_b$ against the exact value. Lower panels plot $L_2$ and $L_\infty$ errors. The dashed lines represent the nominal slopes for the first- and second-order approximations.}
\label{fig:space-convergence}
\end{figure}

We consider in particular the case of a circular pipe of radius $R$, where the solution is known analytically, and is given by the Hagen--Poiseuille parabolic velocity profile: hence, the relationship between the pipe radius $R$, the bulk velocity $U_b$, the fluid kinematic viscosity $\nu$, and the wall friction is known in closed form. Namely, the friction factor $f \equiv 4 \Pi R / \rho U_b^2$ depends on the Reynolds number $Re_b = U_b 2R / \nu$ as $f = 64 / Re_b$. 

We solve numerically equation \eqref{eq:lapl} after discretization on a square domain, of edge length $2.5R$, where a Cartesian mesh with $N \times N$ grid points is defined.  Thanks to the lack of time dependency, the solution can be computed easily by direct matrix inversion.
The improvement provided by the IBM over the staircase approximation is shown in figure \ref{fig:space-convergence}, where the availability of the exact solution is exploited to readily quantify the error. In particular we consider the quantity $f Re_b$, whose exact value is $f Re_b = 64$, its percentage error, and the error in the velocity field evaluated through the norms $L_2$ and $L_\infty$. 
As expected, the staircase approximation is confirmed to only be first-order accurate, whereas the IBM is second-order accurate: with $N=104$, the error in $f Re_b$ with respect to the exact solution is still 3\% for the staircase approximation, whereas it drops to 0.01\% for the IBM. The Python code used for this example is available as Additional Material.

\subsection{The unsteady case}
\label{sec:unsteady}
Since the IBM approach described above acts by adding a correcting weight to the central point of the Laplacian operator, extending it to the Navier--Stokes equations only requires the additional step of considering the time dependence of the solution. 
Let us write the time-dependent, incompressible Navier--Stokes equations after spatial discretisation via second-order central finite-differences on a staggered grid; as an example we take again their $z$ component solved on an $(x,y)$ plane, and emphasize the second derivative stemming from the viscous terms, as:
\begin{equation}
\frac{\dd \wij{i}{j}}{\dd t} =  f_{i,j} = \der{2}{x}{-1} \wij{i-1}{j} + \der{2}{x}{0} \wij{i}{j} + \der{2}{x}{1} \wij{i+1}{j} + \ldots \,   
\label{eq:semidisc-mom}
\end{equation}
where $f_{i,j}$ is a shorthand form for the right-hand side of the discrete equation, which includes the Laplacian involved in the viscous term, as well as the convective and pressure terms. For brevity, in equation \eqref{eq:semidisc-mom} only the second derivative in the $x$-direction is explicitly written, whereas the other contributions to $f_{i,j}$ are grouped in the remainder indicated by ``$\ldots$''.
All the terms at the r.h.s. will be treated with an explicit temporal integration and are thus known from the previous timestep. While the explicit treatment of the right-hand side is not necessary for introducing the present IBM, it is adopted here because it will provide the overall numerical method with interesting properties, which will be discussed in the following. Recall that, even if $f_{i,j}$ is treated explicitly in the momentum predictor equation \eqref{eq:semidisc-mom}, a pressure correction step is still required. 

With reference to the previous example discussed in Sec. \ref{sec:steady}, we assume again that the solid boundary crosses the left arm of the computational stencil, with the body surface lying between the central point $(i,j)$, located within the fluid, and its left neighbor $(i-1,j)$, located within the solid. 
The value $\wij{i-1}{j}$ in equation \eqref{eq:semidisc-mom} is replaced by its linear extrapolation \eqref{eq:interp-lin-extr}, to yield:
\begin{equation}
  \frac{\dd \wij{i}{j}}{\dd t} =  -\lambda_{w;i,j} \wij{i}{j} + \der{2}{x}{0} \wij{i}{j} + \der{2}{x}{1} \wij{i+1}{j} + \ldots \, , 
  \label{eq:semidisc-mom-ibm}
\end{equation}
where $\lambda_{w;i,j}$ is the IBM corrective coefficient arising from the linear extrapolation. The IBM correction has thus removed from the Laplacian the terms involving the neighboring point within the solid, by substituting it with the coefficient $\lambda_{w;i,j}$ that multiplies $\wij{i}{j}$. 
In practice, as long as $\wij{i-1}{j}$ in the solid is zero and the right-hand side is evaluated explicitly, there is no need to modify the discrete Laplacian, since the related term vanishes automatically. In this way the effect of the IBM is to add a term to the right-hand side of equation \eqref{eq:semidisc-mom}, which can thus be rewritten compactly as:
\begin{equation}
  \frac{\dd \wij{i}{j}}{\dd t} =  -\lambda_{w;i,j} \wij{i}{j} + f_{i,j} \, . 
  \label{eq:semidisc-mom-comp}
\end{equation}
If equation \eqref{eq:semidisc-mom-comp} is discretised in time with a fully explicit approach, for instance via an explicit Euler scheme, the well-known explicit extrapolation-based IBMs are obtained. These are known to possess poor stability properties, as they require vanishingly small timesteps whenever the central point $(i,j)$ happens to be very close to the body surface. In this particularly stiff condition, in fact, the ratio $\left(\Delta x - \delta x\right)/\delta x$ contained in $\lambda_{w;i,j}$, tends to infinity.  

Instead, in the following, we keep the explicit treatment of $f_{i,j}$, but allow the IBM term to be treated differently.
The simplest choice, useful for showcasing the method, is to opt for the implicit Euler method, which leads to:
\begin{equation}
  \wij{i}{j}^{*,n+1} = \frac{\wij{i}{j}^{n} + \Delta t f_{i,j}^n }{ 1 + \Delta t  \lambda_{w;i,j}  } \, 
  \label{eq:time-adv-impl}
  \end{equation}
where $\Delta t$ is the time step, and $\wij{i}{j}^{*,n+1}$ is the intermediate velocity of the fractional step method, which needs to be later corrected by an appropriate projection scheme. The implicit treatment of the IBM term has the crucial advantage of not deteriorating (and actually improving) the stability properties of the underlying temporal scheme. This is easily observed by considering the two limiting cases of $\lambda_{w;i,j}=0$, i.e. no IBM correction is applied, and $\lambda_{w;i,j} \rightarrow \infty$, i.e. the point $(i,j)$ is on the body surface. In the first case, we simply recover the unmodified Navier--Stokes equation; in the second, instead, the exact boundary condition $\wij{i}{j}^{*,n+1} = 0$ is enforced. 

Second-order accuracy in time can be achieved without compromising stability by integrating the IBM correction term exactly (more precisely, in a way that would be exact if the r.h.s. were independent of the solution), as explained in the following. 
Let us first consider equation \eqref{eq:semidisc-mom-comp}: this is an ordinary differential equation with a particular solution depending on $f_{i,j}$, assumed here to be constant within a timestep accordingly with the considered explicit temporal scheme, and a homogeneous solution $\tilde{w}_{i,j} = \rme^{-\lambda_{w;i,j} t}$, which can be retrieved by analytical integration for $f_{i,j}=0$. 
In fact, $- \lambda_{w;i,j}$ is the eigenvalue of equation \eqref{eq:semidisc-mom-comp} when $f_{i,j}$ is a constant.
Without loss of generality, for a generic explicit method for temporal integration equation \eqref{eq:semidisc-mom-comp} can be rewritten as follows (dropping superscript $\ast$ and the subscripts to simplify the notation):
\begin{equation}
  A w^{n+1} - B w^{n} =  C F^{n} \,
\label{eq:generic}
\end{equation}
where $F^{n}$ is typically a linear combination of $f_{i,j}$ evaluated at different time levels, as determined by the temporal scheme of choice.
Equation \eqref{eq:generic} reduces to equation \eqref{eq:time-adv-impl} for $A=(1+\lambda_{w;i,j} \Delta t)$, $B=1$ and $C=\Delta t$, with $F^{n} = f_{i,j}^{n}$, when the underlying scheme is a first-order explicit Euler. Alternatively, the corresponding expression for a higher-order (say, Runge--Kutta) scheme is used.

The coefficients $A$, $B$ and $C$ can now be chosen by requesting that equation \eqref{eq:generic} (i) possesses the same eigenvalue as the semi-discrete equation \eqref{eq:semidisc-mom-comp}, and (ii) reduces to the exact steady problem when $\dd w / \dd t = 0$, and thus $w^{n+1} = w^{n}$. 

Constraint (i) can be satisfied by observing that equation \eqref{eq:generic} for $F^{n}=0$ yields $\tilde{w}^{n+1} / \tilde{w}^{n} = B/A$, and by substituting the exact homogenenous solution of equation \eqref{eq:semidisc-mom-comp}, thereby obtaining $B/A = \rme^{-\lambda_{w;i,j} \Delta t}$. 

Constraint (ii) can be enforced by plugging $w^{n+1} = w^{n}$ into equation \eqref{eq:generic}, which yields $w^n \left(A - B\right)/C = F^{n}$, and prescribing that this equation shall equal equation \eqref{eq:semidisc-mom-comp} for $\dd w / \dd t = 0$. This occurs for $\left(A - B\right)/C = \lambda_{w;i,j}$, from which $C=(A-B)/\lambda_{w;i,j}$. The consistency of the underlying temporal scheme already provides $F^{n} = f^{n}$ at steady state.
The three coefficients are now known up to a multiplicative constant, since the problem is linear. By choosing e.g. $C=\Delta t$, one obtains
\begin{equation}
B=\frac{\lambda_{w;i,j} \Delta t}{\rme^{\lambda_{w;i,j} \Delta t} -1}
\label{eq:Bexpr}
\end{equation}
and $A = \lambda_{w;i,j} \Delta t + B$. Thus, equation \eqref{eq:generic} becomes
\begin{equation}
  \left( \lambda_{w;i,j} \Delta t + B \right) w^{n+1} - B w^{n} =  \Delta t F^{n} \, .
\label{eq:expint}
\end{equation}

We note that the function $B(\lambda_{w;i,j} \Delta t)$ of equation \eqref{eq:Bexpr} can in practice be approximated by the reciprocal of a Taylor expansion around $\lambda_{w;i,j} \Delta t=0$ to a desired order ($B=1$ for $1^{\rm st}$ order), without destroying its essential stability property that $1 \ge B \ge 0$ for all $\Delta t$; this may be useful to avoid the evaluation of a transcendental function and a singular limit, and thus improve performance when computing on graphics accelerators.

In the following applications, second-order global temporal accuracy is achieved by adopting the three-stage, third-order Runge--Kutta (RK) method of Ref. \cite{rai-moin-1991} to express $F^n$, owing to its low memory requirements and excellent stability properties which are not affected by the IBM. The RK method is combined with the IBM in the same way it is typically combined with Crank-Nicolson for the implicit integration of the viscous term, as in \cite{kim-moin-moser-1987}.  This choice results in the following scheme:
\begin{equation}
  \left( \lambda_{w;i,j} c_k \Delta t + B \right) \wij{i}{j}^{\ast,n+\frac{k}{3}} - B \wij{i}{j}^{n+\frac{k-1}{3}} =  \Delta t  \left[ a_k f_{i,j}^{n+\frac{k-1}{3}} + b_k f_{i,j}^{n+\frac{k-2}{3}} \right] \, \quad \text{for} \,\, k=1,\ldots,3 , 
  \label{eq:rk3-rai}
\end{equation} 
where $\wij{i}{j}^{n+\frac{k}{3}}$ is the velocity at the $k$-th RK stage, whereas $f_{i,j}^{n+\frac{k}{3}}$ is the right-hand side evaluated with $\wij{i}{j}^{n+\frac{k}{3}}$ and other variables at the same RK stage. We also recall that the intermediate velocity (denoted via the additional superscript $\ast$) is not divergence-free and an additional projection step is required after each substage to obtain the solenoidal velocity field. According to \cite{rai-moin-1991}, the coefficients in equation \eqref{eq:rk3-rai} are: 
\begin{equation}
  a_k = \left\{\frac{64}{120}, \frac{50}{120}, \frac{90}{120}\right\}; \qquad b_k = \left\{0, - \frac{34}{120}, - \frac{50}{120}\right\}; \qquad c_k = a_k + b_k .
\end{equation}
 



\subsection{The unsteady case: example}
\label{sec:example-unsteady}

The simple example of the laminar parallel flow in a circular pipe of radius $R$, already considered in Sec. \ref{sec:example-steady}, is extended here to demonstrate the present IBM in a time-dependent flow. By discretizing the problem on a fine Cartesian mesh with $200 \times 200$ grid points, we ensure that the spatial discretization error is not dominant. The flow is governed by the unsteady version of equation \eqref{eq:lapl}, i.e.
\begin{equation}
\frac{\partial w}{\partial t} = -\Pi + \frac{1}{Re} \nabla^2 w \, ,
\label{eq:unstlapl}
\end{equation}
where the imposed uniform pressure gradient $\Pi (t)$ can now vary with time. We opt for the time dependency $\Pi (t) = \sin (2\pi f t)$, where frequency $f$ enables the definition of a Reynolds number $Re=R^2 f / \nu$, i.e. the ratio between the characteristic diffusion time $R^2/\nu$ and the characteristic time of the forcing $1/f$. 

For a given temporal integration scheme, such as the Runge--Kutta method described above, the accuracy of the numerical solution of equation \eqref{eq:unstlapl} depends on how well the temporal evolution of $\Pi$ and of the diffusive effects are represented. This is quantified by two nondimensional numbers: the nondimensional timestep $f \Delta t$, and the grid P\'eclet number $Pe = \nu \Delta t/ \Delta x^2$. By selecting $Re=50$ we make sure that even for the lower values of $\Delta t$ tested below, the main source of error will not be the temporal evolution of $\Pi$ but rather the temporal accuracy of the viscous effects and of the IBM method. 

\begin{figure}
\centering
\includegraphics[]{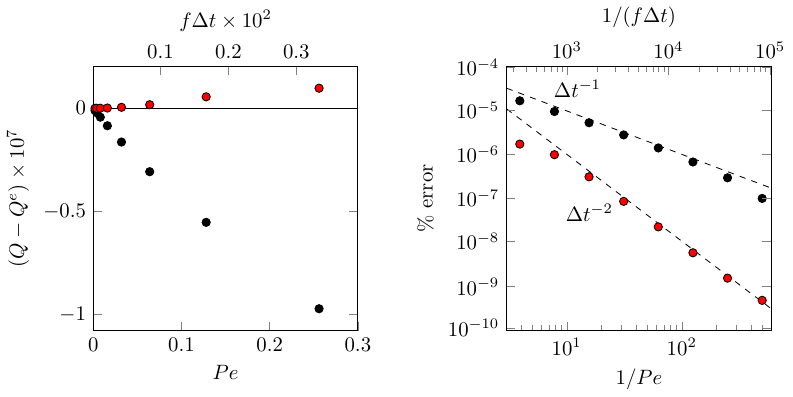}
\caption{Unsteady laminar pipe flow, equation \eqref{eq:unstlapl}, spatially discretized on a square domain with $200 \times 200$ grid points. Left: evolution of the difference $Q - Q^{(e)}$ with the temporal resolution, expressed via the P\'eclet number $Pe = \nu \Delta t/ \Delta x^2$; the asymptotic flow rate value is obtained via Richardson extrapolation. Right: percentage error $100 |Q-Q^{(e)}|/Q^{(e)}$. The dashed lines represent the nominal slopes for the first- and second-order temporal accuracy.}
\label{fig:time-convergence}
\end{figure}

Starting from an initial condition of quiescent flow, equation \eqref{eq:unstlapl} is integrated in time up to $t=0.5/f$, and at that time the flow rate $Q$ in the pipe is measured; the numerical experiment is repeated for several values of $f \Delta t$ ranging between $0.002$ and $1.5625 \times 10^{-5}$. 
Note that the largest value of the time step corresponds, for the employed spatial discretization, to $Pe=0.256$, which is less than half the critical P\'eclet number for the temporal stability of the underlying RK method. 
Equation \eqref{eq:unstlapl} is solved by the second-order method presented in equation \eqref{eq:rk3-rai}, as well as by the first-order method obtained from the combination of the RK method with an implicit Euler scheme for the IBM term, i.e. $A=1+\lambda c_k \Delta t$, $B=1$ and $C=\Delta t$. 
Results are displayed in figure \ref{fig:time-convergence}, together with the error with respect to the estimate of exact flow rate $Q^{(e)}$, obtained via Richardson extrapolation. The figure shows both that the expected  order of temporal convergence is achieved and that the error is generally extremely low. 

As a final remark, it is interesting to note that a classic explicit treatment of the IBM term yields a stable integration only for the smallest timestep tested above. The poor stability of the explicit IBM could be even worse if any grid point happens to be very close to the body surface, making the method unconditionally unstable. Crucially, the present implicit treatment removes this issue, and averts deteriorating the stability properties of the underlying numerical method.

%% file: discussion.tex
\section{Discussion}
\label{sec:discussion}

\subsection{The near-wall dominance of viscous terms} 
The assumption that, near the wall, the viscous terms dominate over the non-linear and pressure terms, and that, as a consequence, a correction to the Laplacian is sufficient for the IBM to work correctly to yield a second-order method, is a crucial step that deserves some additional comments.

While the correctness of the procedure has been demonstrated in the examples above, and will be reinforced by the applications described below, it is noteworthy that the physics-based assumption of viscous terms dominance near the wall \cite{orlandi-leonardi-2006} is equivalent to the assumption that, near the wall, the velocity profile is linear. It is in this regard that the IBM is said to suit direct numerical simulations of turbulence (and possibly wall-resolved large-eddy simulations), which require a grid step small enough to represent well this linear region.
Since any velocity profile is linear when observed close enough to the wall, the requirement that the Laplacian correction is sufficient to fully achieve second-order accuracy superposes with the step requirements of a direct numerical simulation.

Another aspect whose discussion has been omitted is moving boundaries. 
We have given most attention to stationary boundaries because this is the context of our practical applications of the method. 
An additional use case where an IBM potentially offers notable advantages is when the flow around moving bodies is considered \cite{kim-choi-2019}. 
The present IBM can be applied straightforwardly to the case of moving rigid bodies. In fact, the method works as long as the velocity profile near the boundary is linear, and this remains the case also when the body is moving, provided one applies the IBM correction in a reference frame in which the body is stationary. This is readily accomplished by subtracting the velocity of the rigid body from the local fluid velocity.

\subsection{The implicit nature of the immersed-boundary correction}

The greatest novelty of the approach outlined above consists in the implicit nature, in both space and time, of the correction introduced to account for the presence of the boundary; the ability to concentrate the correction in the central point of the stencil allows this feature to be achieved at no cost, in a way that extends readily to the multi-dimensional case.

In the stationary case described in Sec. \ref{sec:steady}, the term ``implicit'' refers to the fact that the value $\wij{i-1}{j}$ in a point within the body is not actually computed by using equation \eqref{eq:interp-lin-extr}, but gets hard-coded instead into the expression of the discretised Laplacian and in particular into $\tilde{d}^{(2)}_{x;ij}(0)$ through the coefficient $\lambda_{w;i,j}$.
Eliminating the extrapolated values entails essentially no computational cost,  and makes the discretization matrix more diagonally dominant, see equation \eqref{eq:disclapl-extr}, so that the convergence of any iterative method used to solve the discretised equations is improved rather than deteriorated. 
This property becomes apparent when one considers that $d^{(2)}_{x;ij}(-1)$ and $d^{(2)}_{x;ij}(0)$ are of opposite signs, and thus $\tilde{d}^{(2)}_{x;ij}(0)$ monotonically increases for $\Delta x \ge \delta x \ge 0$ and is prevented from approaching $0$. This cures the numerical stability problems that plague ``explicit'' methods when the boundary point approaches a discretisation node (i.e. $\delta x \rightarrow 0$). Such formulations imply that $\wij{i-1}{j} \rightarrow \infty$, whereas the implicit correction for $\delta x \rightarrow 0$ makes the denominator of equation \eqref{eq:time-adv-impl} tend to infinity, and the value of $\wij{i}{j}$ gradually approach zero, which is the desired result. 

In the time-dependent case of Sec. \ref{sec:unsteady}, and within our design choice of using an explicit time integration scheme, ``implicit'' additionally means that the linear extrapolation of equation \eqref{eq:interp-lin-extr} is evaluated at time $t + \Delta t$, as in equation \eqref{eq:time-adv-impl}, to preserve the convergence properties of the steady case. This again avoids numerical instabilities:
for $\delta x = 0$, equation \eqref{eq:time-adv-impl} can be shown to equal the exact implicit boundary condition $\wij{i}{j}^{*,n+1} = 0$.
%

Overall, the implicit correction of the IBM ensures convergence and stability of the numerical method; moreover, since only the central point of the Laplacian stencil is modified, this improvement is obtained at no computational and memory cost, since the velocity $\wij{i-1}{j}$ at the ghost point within the solid region needs to be neither explicitly computed nor explicitly stored.
This additional advantage obviates another programming difficulty related to the presence of external points appearing in more than one equation (for example the Cartesian components of the momentum equation); in the explicit implementation multiple values extrapolated linearly from different directions would have to be stored (or otherwise a higher-than-linear extrapolation would be required), but no such difficulty arises with the implicit formulation.


\subsection{The underlying staircase approximation}

As pointed out in Sec. \ref{sec:steady}, the simplest (first-order) description of the immersed boundary is achieved by a staircase (piecewise-constant) approximation of the body geometry, in which the boundary always coincides with a grid point.
Since the IBM uses a linear extrapolation to improve upon the staircase approximation of the boundary and to restore the original second-order spatial accuracy of the numerical method, it is essential that the underlying staircase approximation works properly, before any correction is applied. This involves non-trivial aspects.

To begin with, the equations of motion in their discrete form need to be closed, i.e. the number of unknowns must equal the number of equations. This property becomes non-obvious when the geometry is complex and the grid is staggered. To fulfill the closure requirement, the discretization grid is defined independently for each velocity component, and each grid is compared with the true boundary to tag a grid point as either ``internal'' (inside the fluid region) or ``external'' (inside the solid region).
For each component, internal points are chosen as those where the corresponding component of the momentum equation \eqref{eq:semidisc-mom} needs to be solved; this ensures the correspondence between equations and unknowns. The velocity components on the external points are set to zero. 
This decision, which is trivial in the one-dimensional case, avoids any trouble with external points appearing in more than one equation.

The closure of the continuity equation deserves a specific remark. Since a pressure boundary condition is neither required nor present, only the velocity grids are compared to the boundary to discriminate external and internal points. Thereafter, a pressure point is labeled ``internal'' if it falls on either side of at least one internal velocity point; it follows that an ``internal'' pressure point may occasionally fall (slightly) outside the true boundary (in the solid region), and may also be shared by different components of the momentum equation in different directions. 
To match the number of equations and unknowns, the continuity equation is then in principle solved everywhere, both in the internal and external pressure collocation points, as though there was no boundary. In practice, however, only the internal points need to be considered since, with this definition, external pressure points are surrounded by all zero velocities, and continuity is trivially satisfied there. In practice, when the pressure-correction step of the fractional step method is executed, only the internal pressure collocation points are updated, and these are the only ones that will appear in the momentum equations at the following time step. 

\subsection{Neumann-type boundary conditions}

So far, we have introduced and discussed the IBM by focusing on Dirichlet-type boundary conditions. This covers the majority of use cases in fluid mechanics, starting from the widespread no-slip and no-penetration condition at a solid wall.

However, Neumann-type boundary conditions may be required. Our IBM does not preclude the use of Neumann-type boundary conditions. However, since such generality causes additional complications, for maximum simplicity the method has been presented above in a form where Neumann-type boundary conditions can be handled only for cases where the IBM is not directly involved. These still constitute a large fraction of the cases of interest in fluid mechanics. 
For example, a Neumann condition is used at inlet/outlet surfaces, which are typically not pre-determined and can thus be easily made to consist in planes parallel to coordinate axes. A further situation where Neumann conditions are easily dealt with is when free surfaces exist which are flat because of gravity forces. 
Also, the method works without modifications for partial slip (ensuing from equivalent boundary conditions for riblets, super-hydrophobic surfaces, or other means for skin-friction drag reduction), where the first derivative is multiplied by a coefficient of the order of the grid step: this can be understood either because these are equivalent to a Dirichlet condition applied on a (displaced) plane surface, or because a first-order derivative multiplied by a first-order coefficient yields a second-order condition.

%% file: results.tex
\section{Results}
\label{sec:results}

We move on to describe three application examples, in which the IBM with the convective terms and the continuity equation are at play. The first is a classic, simple two-dimensional flow, useful to confirm the correctness of the solution through comparison with literature data. The remaining two involve turbulent flows over non-planar boundaries. One is relatively simple from the geometric standpoint, and concerns the turbulent flow in an open channel with a sinusoidally undulated bottom. The other case is about the flow within the human nose, and is instead characterized by an extremely complex three-dimensional geometry.

\subsection{The laminar flow around a circular cylinder}
We start with the classic, relatively simple two-dimensional laminar flow around a circular cylinder of diameter $D$ immersed in a uniform stream $U_\infty$. Reference data are abundant in literature, and can be reproduced easily as long as the value of the Reynolds number $Re = U_\infty D / \nu$ remains limited. The same test case was considered in Ref.\cite{pinelli-etal-2010}, which we replicate here. We consider the same two cases: $Re=30$, where the flow is steady, and $Re=185$, where the flow is unsteady and exhibits periodic vortex shedding.
The dimensions of the computational domain are mutuated from Ref.\cite{pinelli-etal-2010} : the cylinder is placed at the origin of the coordinate axes, and the domain has streamwise extent of $[-9D, 40D]$ and vertical extent $[-17D,17D]$. 
In a region surrounding the cylinder, the grid is uniform with a step of $0.02 D$, as in one of the cases in \cite{pinelli-etal-2010}.

\begin{table}
\centering
\begin{tabular}{l|ccccc}
  & $C_d$ &  $l/D$ & $a/D$ &  $b/D$ & $\theta$ [$\deg$]\\
\hline
  present                      & $1.819$ &  $1.5837$ & $0.5410$ & $0.5296$ & $51.8$ \\
  freeFEM                      & $1.784$  &  $1.6293$ & $0.5512$ & $0.5311$ & $47.4$ \\
  Ref.\cite{pinelli-etal-2010} & $1.80$   &  $1.70$   & $0.56$   & $0.52$   & $48.5$ \\
  Ref.\cite{tritton-1959}      & $1.75$   &       -   & -        & -        & -      \\
  Ref.\cite{coutanceau-bouard-1977} & -   &  $1.55$   & $0.54$   & $0.54$   & $50.0$ \\
\end{tabular}
\caption{Quantitative features of the cylinder flow at $Re=30$: drag coefficient $C_d$, length $l$ of the separated region, streamwise distance $a$ of the center of wake vortices, vertical separation $b$ of the center of wake vortices, angular position $\theta$ of the separation point.}
\label{tab:re30}
\end{table}

\begin{figure}
\centering
\includegraphics[width=0.49\textwidth]{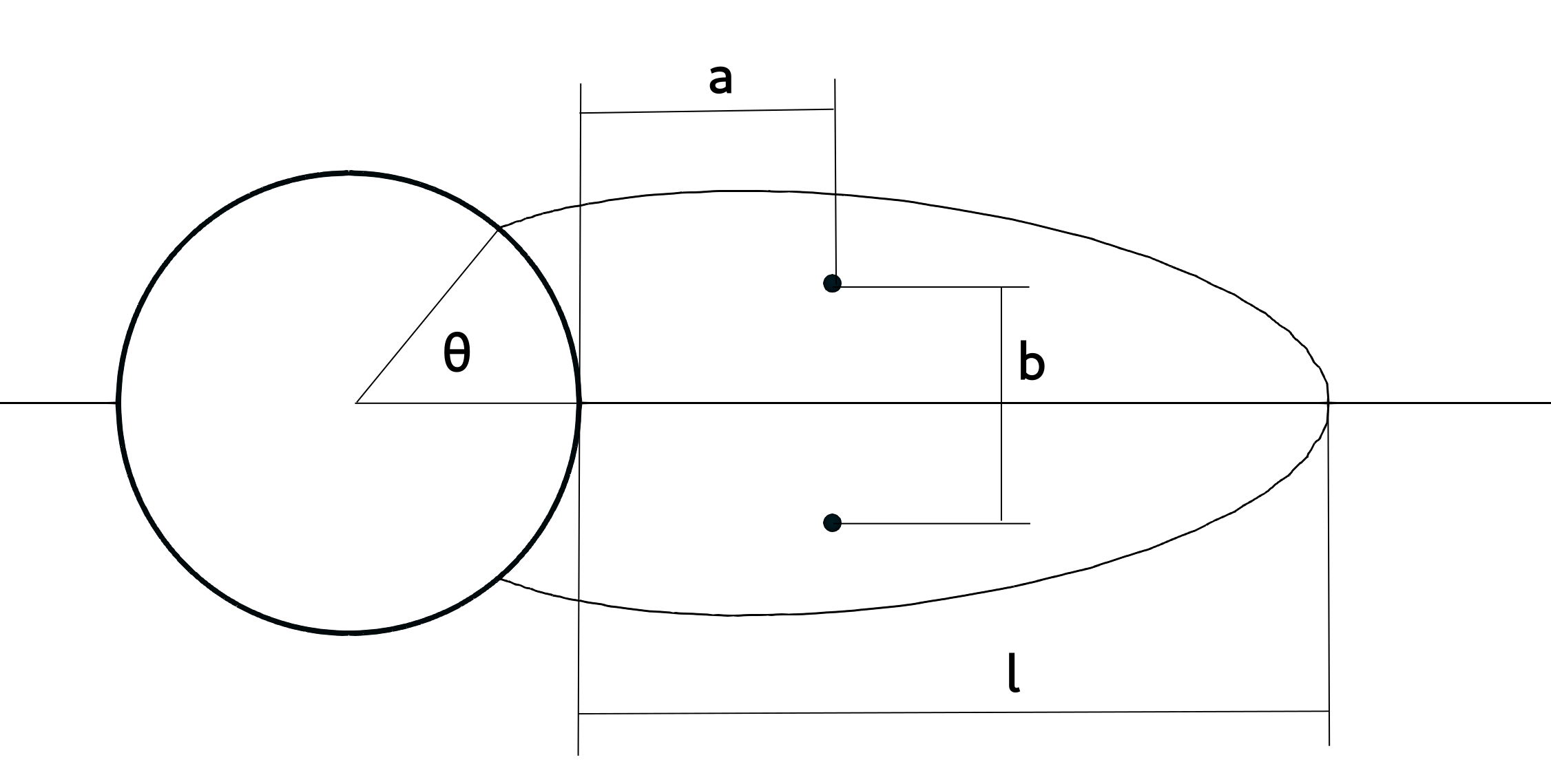}
\includegraphics[width=0.49\textwidth]{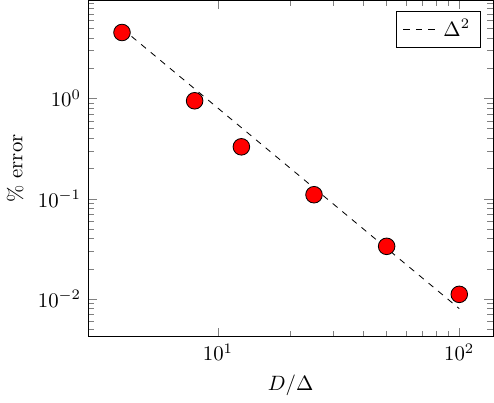}
\caption{Left: sketch of the cylinder flow, with the definition of the parameters $l$ (length of the recirculation), $a$ (horizontal position of the center of the wake vortices), $b$ (vertical distance between the centers of the two wake vortices) and $\theta$ (angular position of the separation point). Right: percentage error of the total drag force versus spatial resolution. The dashed line shows the expected second-order decrease.}
\label{fig:cylinder-sketch}
\end{figure}

The drag of the cylinder and the parameters of the wake resulting from the simulation are in very good quantitative agreement with available data. Table \ref{tab:re30} reports the drag coefficient, the length of the recirculating region, the horizontal and vertical distance between the centers of the two wake vortices, and the angular position of the separation point. 
The meaning of the wake parameters is sketched in figure \ref{fig:cylinder-sketch} (left).
These values are compared with information from Ref.\cite{pinelli-etal-2010} and references therein, as well as with the results of a companion simulation carried out with the non-commercial software FreeFEM \citep{hecht-2012}, which solves the two-dimensional, steady Navier–Stokes equations using the Newton algorithm with a spatial discretisation that uses quadratic elements (P2) for the velocity and linear elements (P1) for the pressure, with discretization on 86752 triangles. 
Another test (in the slightly different configuration of cylinder between two walls, useful to shrink the computational domain) has measured the decrease rate of the error in the total drag force. The plot in figure \ref{fig:cylinder-sketch} (right) demonstrates once again the second-order accuracy of the solution.

\begin{figure}
\centering
\includegraphics[width=\textwidth]{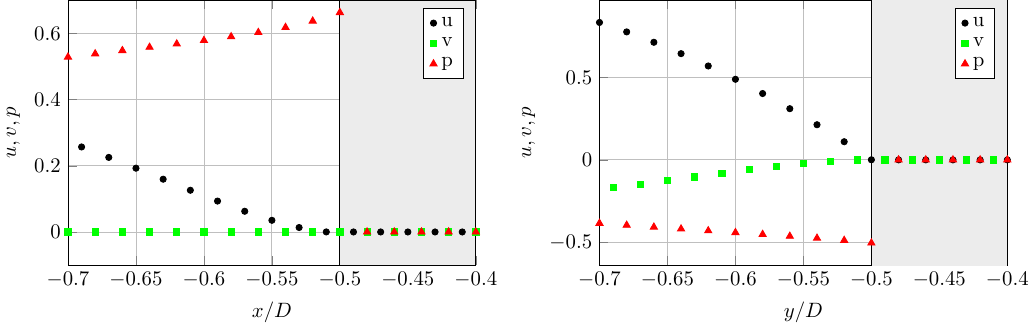}
\caption{Velocity and pressure profiles near the wall of the circular cylinder at $Re=30$, at the stagnation point $(-D/2,0)$ as a function of $x/D$ (left), and at the bottom side of the cylinder $(0,-D/2)$ as a function of $y/D$ (right). The shaded area marks the interior of the body.}
\label{fig:cylinder-zoom}
\end{figure}

Figure \ref{fig:cylinder-zoom} shows the local solution in the proximity of the stagnation point at $x/D=-0.5, y/D=0$ (left panel), and near the bottom side of the cylinder at $x/D=0$ and $y/D=-0.5$. The grey area indicates the range of coordinates corresponding to the cylinder interior. The solution, in terms of both velocity components and pressure, is extremely regular in the proximity of the stagnation point, no oscillations are visible and the no-slip boundary condition is satisfied. This is remarkable also on consideration of the relatively large size of the grid, which would emphasize any inaccuracy near the boundary.

\begin{table}
\centering
\begin{tabular}{l|ccc}
  & $C_d$ &  $C_l'$ & $St$\\
\hline
  present       & $1.382$   &   $0.470$ & $0.198$ \\  
  freeFEM       & $1.365$   &   $0.462$ & $0.196$ \\
  Ref.\cite{pinelli-etal-2010}& $1.430$     & $0.423$ & $0.199$ \\
  Ref.\cite{vanella-balaras-2009}    & $1.377$  &   $0.428$ &  - \\
  Ref.\cite{guilmineau-queteuy-2002} & $1.280$  &   $0.443$ &  $0.195$ \\
  Ref.\cite{lu-dalton-1996}         & $1.310$  &   $0.443$ &  $0.195$ \\
  Ref.\cite{williamson-1988} & -        &   - &  $0.193$ \\
\end{tabular}
\caption{Quantitative features of the cylinder flow at $Re=185$: drag coefficient $C_d$, r.m.s. value of the lift coefficient $C_l'$, and Strouhal number $St$.}
\label{tab:re185}
\end{table}

The cylinder flow at the higher $Re=185$ is unsteady and periodic, and therefore is home to a richer physics, observed through a time-dependent lift and a periodic shedding with a well defined frequency $f$. These are quantified through the root-mean-square value of the lift coefficient $C_l'$ and the value of the Strouhal number $S= f D / U_\infty$. Values of $C_l'$ and $S$ obtained with the present method are reported in table \ref{tab:re185}, together with the drag coefficient that, as expected, lowers from the value at $Re=30$. Except for the value of $Re$, the simulation replicates the one described above, and is run for an integration time sufficient to cover 50 shedding periods. Even in this case, a companion simulation with identical parameters run with FreeFEM is presented; it uses the Uzawa algorithm and the Adams--Bashfort scheme for time integration, with a dimensionless time step of 0.002. 
The comparison with available literature data, including those from Ref.\cite{pinelli-etal-2010}, confirms once again the ability of the present method to quantify correctly the flow features of interest.

\subsection{The turbulent flow in a channel with undulated bottom}
\label{sec:wavy}

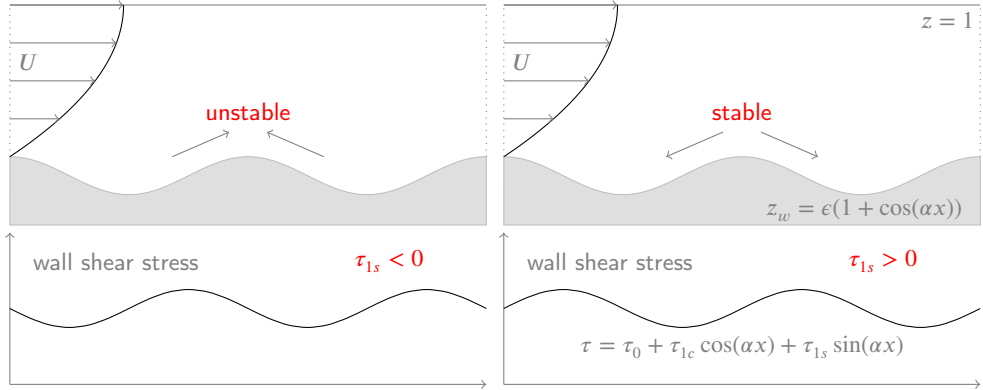
\begin{figure}
\centering
\input{wavybottom-unstable.tex}
\input{wavybottom-stable.tex}
\caption{Effect of the wall shear stress on the stability of a sand bed. The flow develops between the wavy wall described by $z_w(x)$ and the upper boundary at $z=1$. The wall-shear stress is decomposed into in-phase and quadrature components. Left: unstable situation (the quadrature component $\tau_{1s}$ of the stress is negative). Right: stable situation ($\tau_{1s}$ is positive). Adapted from Ref. \cite{luchini-2016}.}
\label{fig:wavy-sketch}
\end{figure}

The IBM is now applied to simulate via DNS the turbulent flow in a channel with a sinusoidally undulated bottom wall: this is an idealised model of a river or flume with a sandy bottom, that may bulge up and generate ripples and dunes.
The generation mechanism of these ripples involves fluid inertia \cite{charru-etal-2013}; a net accrual or depletion of sand particles occurs depending on the relative spatial phase between the fluid shear stress and the wall undulation. 
Figure \ref{fig:wavy-sketch} sketches a sinusoidally undulated bottom and the relative wall-shear stress, decomposed into in-phase (cosine) and quadrature (sine) components, whose sign determines the generation of ripples \cite{blondeaux-1990}: they grow whenever an unstable situation is determined by a negative quadrature component of the wall-shear stress. 

We replicate with higher accuracy the simulations performed by Luchini in Ref. \cite{luchini-2016}, aimed at determining how this instability depends on the wavelength and elevation of the bottom.
We consider a channel configuration with a shear-free, flat upper surface, with lengths and velocities made dimensionless with the channel height $h$ and the bulk velocity $U_b$. The (dimensionless) expression of the undulated bottom is:
\[
z_w = \epsilon \left( 1 + \cos \alpha x \right).
\]
Here $\alpha$ is the dimensionless wave number of the wavy bottom wall (with wave length $\lambda = 2 \pi / \alpha$), and $\epsilon$ is the dimensionless amplitude of the waviness. The computational domain extends for $L_x= \lambda$, $L_y=2\pi$ and $L_z=1$ in the streamwise, spanwise and wall-normal directions. The wavenumber is varied in the range $1/256 \le \alpha \le 1/2$ (or $12.5 \le L_x \le 1608$). Please notice that these very long computational boxes are required because this flow was shown in \cite{luchini-2016} to have an interesting response to very long wavelengths of the undulation. While the shortest extreme is comparable to that employed in low-$Re$ turbulent channel flow simulations \cite{kim-moin-moser-1987}, at the other extreme more than two-orders of magnitude longer domains are considered.
Several simulations are run for the Reynolds number $Re_b = U_b h /\nu = 2800$. Periodic boundary conditions are used for the streamwise and spanwise directions, free-slip boundary conditions are applied at $z=1$, and no-penetration and no-slip conditions are applied at the wavy wall. 
To obtain well converged statistics, simulations are run for at least $1000 U_b/h$ time units after the initial transient. Table \ref{tab:wavy-parameters} contains general information about the set of simulations carried out for the present work. In a first set of simulations, the grid spacing is kept constant while the size of the computational domain is varied; a second set of simulations, used for a grid convergence study, has a fixed domain size, and a spatial resolution that progressively increases uniformly in all directions. Since the wall-normal distribution of grid points is mildly stretched, an average wall-normal spacing $\Delta z = 1/N_z$ is used to provide an equivalent uniform grid spacing $\Delta = ( \Delta x \Delta y \Delta z)^{(1/3)}$.

\begin{table}
\centering
\begin{tabular}{ccrcccccccc}
$\epsilon$ &  $\alpha$ & $N_{tot}$ &  $\Delta x$ & $\Delta y$ & $\Delta z_{min}$ &  $\Delta z_{max}$ &  $N_z$ & $\Delta$ & $\Delta t$ \\
\hline
$0.05$     &  $1/256\le \alpha \le 1/2$ & up to $251657210$ & $0.0654$ & $0.0393$ & $4.0\times 10^{-3}$ & $0.0224$ & $64$  & $0.034245$ & $0.03$ \\
$0.025$    &  $1/128\le \alpha \le 1/2$ & up to $251657210$ & $0.0654$ & $0.0393$ & $2.0\times 10^{-3}$ & $0.0111$ & $128$ & $0.027180$ & $0.008$ \\
$0.0125$   &  $1/32 \le \alpha \le 1/2$ & up to $125828605$ &  $0.0654$ & $0.0393$ & $9.6\times 10^{-4}$ & $0.0056$ & $256$ & $0.021573$ & $0.0015$ \\
$0.05$     &  $1/2$                     & $1966080$ &  $0.0654$ & $0.0393$ & $4.0\times 10^{-3}$ & $0.0225$ & $64$  & $0.034245$ & $0.001$ \\
$0.05$     &  $1/2$                     & $2799360$ &  $0.0582$ & $0.0349$ & $3.6\times 10^{-3}$ & $0.0199$ & $72$  & $0.030442$ & $0.001$ \\
$0.05$     &  $1/2$                     & $3840000$ &  $0.0524$ & $0.0314$ & $3.2\times 10^{-3}$ & $0.0179$ & $80$  & $0.027398$ & $0.001$ \\
$0.05$     &  $1/2$                     & $6635520$ &  $0.0436$ & $0.0262$ & $2.6\times 10^{-3}$ & $0.0149$ & $96$  & $0.022830$ & $0.001$ \\
$0.05$     &  $1/2$                     & $8436480$ &  $0.0403$ & $0.0242$ & $2.4\times 10^{-3}$ & $0.0137$ & $104$ & $0.021088$ & $0.001$ \\
$0.05$     &  $1/2$                     & $10536960$ &  $0.0374$ & $0.0224$ & $2.2\times 10^{-3}$ & $0.0127$ & $112$ & $0.019557$ & $0.001$ \\
$0.05$     &  $1/2$                     & $15728640$ &  $0.0327$ & $0.0196$ & $2.2\times 10^{-3}$ & $0.0111$ & $128$ & $0.017108$ & $0.001$ \\
$0.05$     &  $1/2$                     & $30720000$ &  $0.0262$ & $0.0157$ & $1.6\times 10^{-3}$ & $0.0089$ & $160$ & $0.013699$ & $0.001$ \\
\end{tabular}
\caption{Computational parameters for the numerical simulations of the turbulent flow in a channel with undulated bottom. The equivalent grid spacing $\Delta$ is computed by assuming a uniform grid in all directions.}
\label{tab:wavy-parameters}
\end{table}
 
The quantities of interest here are the spatial Fourier components $\tau_{1c}$ and $\tau_{1s}$ of the time-averaged wall-shear stress $\tau(x)$. They are defined as
\begin{equation}
\tau_{1c} = \frac{2}{L_x} \int_0^{L_x} \tau(x) \cos \left( \frac{2 \pi x}{L_x} \right) \text{d} x; \qquad
\tau_{1s} = \frac{2}{L_x} \int_0^{L_x} \tau(x) \sin \left( \frac{2 \pi x}{L_x} \right) \text{d} x.
\end{equation}

\begin{figure}
\centering
\includegraphics[width=0.49\textwidth]{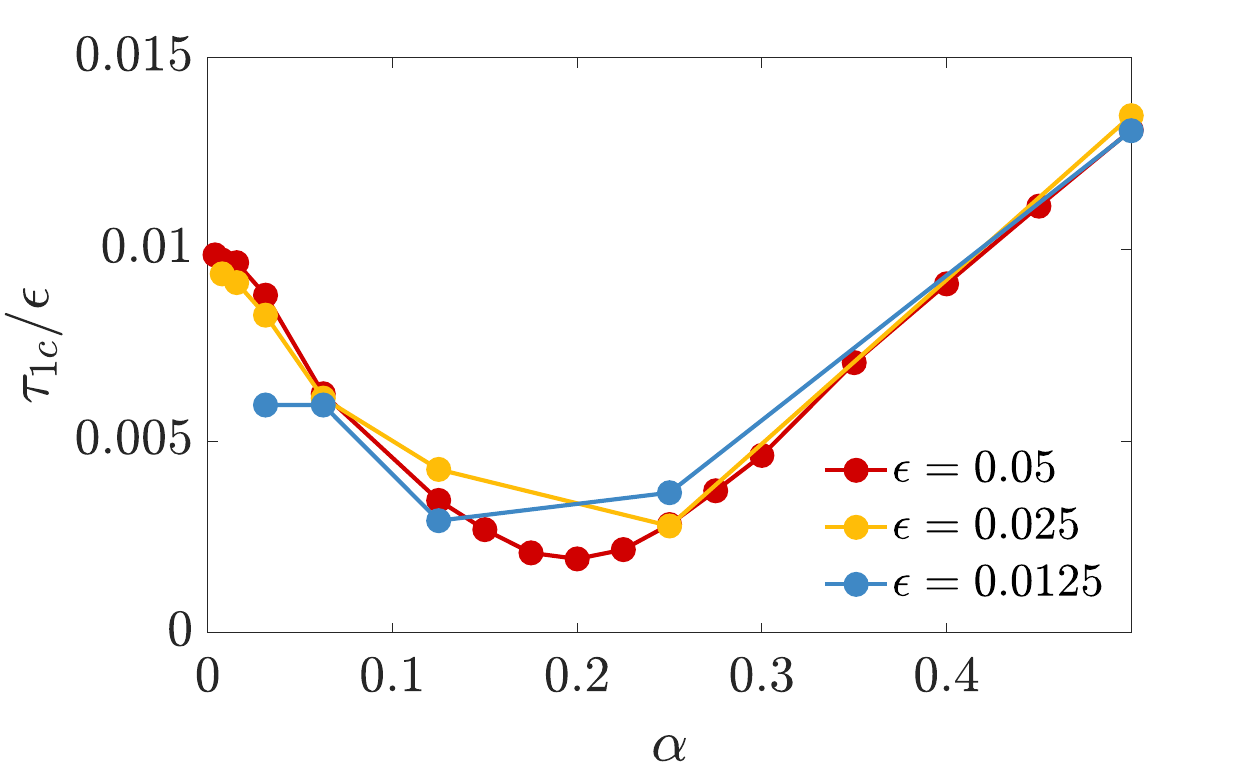}
\includegraphics[width=0.49\textwidth]{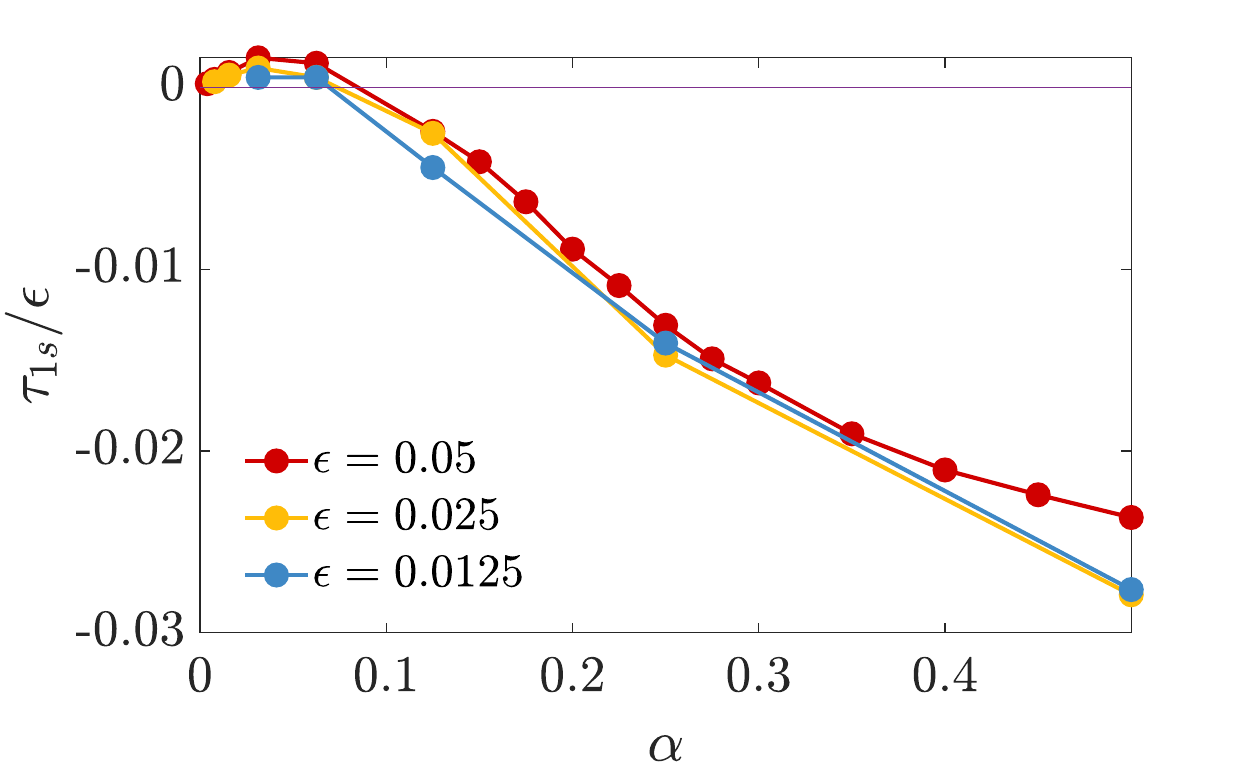}
\caption{In-phase (left) and quadrature (right) components $\tau_{1c}$ and $\tau_{1s}$ of the wall-shear-stress response to a wall modulation of wavenumber $\alpha$.}
\label{fig:wavy-results}
\end{figure}

\begin{table}
\centering
\begin{tabular}{rrrrrr}
\hline
$N_{tot}$  &  $\Delta$ & $\tau_{1c} \times 10^4$ &   $\sigma_{1c} \times 10^6$     & $\tau_{1s} \times 10^3$ &  $\sigma_{1s} \times 10^6$ \\ 
1966080    &  0.034245 	 &  6.40764  &   7.18508  &   -1.18162   &     6.40058 \\
2799360    &  0.030442	 &  6.76041  &   8.67627  &   -1.25258   &     7.84068 \\
3840000    &  0.027398	 &  6.80707  &   4.46546  &   -1.29628   &     3.54087 \\
6635520    &  0.022830	 &  7.13429  &   7.65407  &   -1.37018   &     6.87870 \\
8436480    &  0.021088	 &  7.28166  &   7.05332  &   -1.38633   &     6.14717 \\
10536960   &  0.019557	 &  7.28948  &   8.31248  &   -1.40233   &     6.93155 \\
15728640   &  0.017108	 &  7.42786  &   8.55187  &   -1.42513   &     7.48778 \\
30720000   &  0.013699	 &  7.46693  &   7.73307  &   -1.45979   &     7.64443 \\
\end{tabular}
\caption{Spatial convergence study for the wavy channel test case. The table lists the number of grid points, the mean grid spacing, the values $\tau_{1c}$ and $\tau_{1s}$ of the mean stresses, and the corresponding root-mean-square values $\sigma$ of the variance of the estimate of the mean, computed after \cite{russo-luchini-2017}.}
\label{tab:wavy-convergence}
\end{table}

Figure \ref{fig:wavy-results} shows the dependence of $\tau_{1c}$ and $\tau_{1s}$ on the wavenumber $\alpha$ of the wall modulation, for different values of $\epsilon$. The red curve is for $\epsilon=0.05$ and has been obtained using the same time step and grid size as in \cite{luchini-2016}, whose results are perfectly reproduced. 
The plot confirms that the considered values of $\epsilon$ are small enough to be in the linear regime.
As already observed by \cite{luchini-2016}, when $\alpha$ is reduced $\tau_{1c}$ decreases to a local minimum at $\alpha \approx 0.2$, and then increases again. At the smallest $\alpha$, the quadrature component $\tau_{1s}$ changes sign, indicating a change of the stability of the ripples. This is confirmed at different amplitudes $\epsilon$. For a discussion of the physical implications of this behaviour, see \cite{luchini-2016}.

\begin{figure}
\centering
\includegraphics[width=0.8\textwidth]{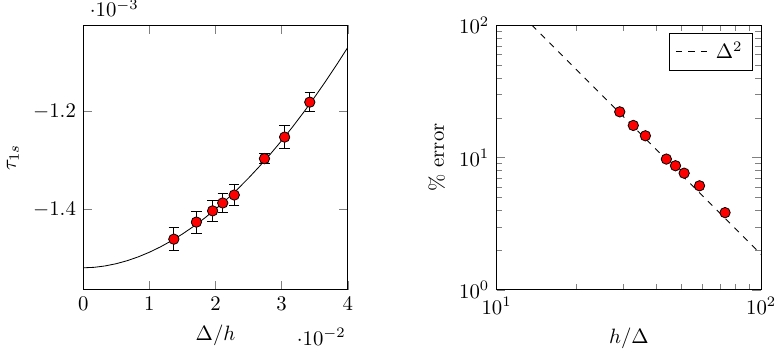}
\caption{Convergence of the quadrature component $\tau_{1s}$ of the wall shear stress for the wavy channel test case. Left: variation of the time-averaged value of $\tau_{1s}$ with the mean spatial resolution $\Delta$; the error bars correspond to $\pm 3 \sigma_{1s}$, where the root-mean-square value $\sigma_{1s}$ of the estimate of the mean is computed after \cite{russo-luchini-2017}. The continuous line is a fit of equation \eqref{eq:conv}, yielding $\tau_{1s}^{(e)}=-0.001518$, $C=0.2109$ and $p=1.912$. Right: percentage error $100 (\tau_{1s}^{(e)}-\tau_{1s})/\tau_{1s}^{(e)}$ of the quadrature stress versus spatial resolution; the dashed line shows the expected second-order decrease.}
\label{fig:wavy-convergence}
\end{figure}

To verify the order of accuracy of the IBM, figure \ref{fig:wavy-convergence} plots the results of a spatial convergence study, whose main results are reported in table \ref{tab:wavy-convergence}. For $\alpha=0.5$ (or $L_x=4\pi$) and $\epsilon=0.05$, different discretizations have been tested, by progressively reducing the mesh size uniformly in all the three directions. 
The coarsest mesh has $(N_x,N_y,N_z)=(192,160,64)$, corresponding to a total number of points $N_{tot} \approx 2 \times 10^6$. 
The finest mesh has $(N_x,N_y,N_z)=(480,400,160)$, with a total number of $ \approx 3 \times 10^7$ points. The time step is fixed for all cases at $\Delta t = 0.001$, a value that is small enough to ensure that the time discretisation error is not dominant. 

Given a numerical method with rate of convergence $p$, the difference between the solution $f$ computed on a three-dimensional uniform grid with spacing $\Delta$ and the exact solution $f^{(e)}$ must vary according to 
\begin{equation}
|| f - f^{(e)} || \le C \Delta^{-p}
\label{eq:conv}
\end{equation} 
where $C$ is a constant. We use equation \eqref{eq:conv} to compute a least-square fit for the quadrature stress component (equivalent results, not shown, are obtained with the in-phase component) obtained from the numerical experiments. Since the exact solution is not known, the value $\tau_{1s}^{(e)}$ is evaluated with the asymptotic value of the fitting curve. 
In computing the time-averaged value of $\tau_{1s}$, we also quantify the error implied by the finite averaging time; to this purpose the method described by Russo and Luchini in \cite{russo-luchini-2017} is used to compute the root-mean-square value $\sigma$ of the estimate of the mean, and figure \ref{fig:wavy-convergence} plots error bars for $\pm 3 \sigma$.
As expected, figure \ref{fig:wavy-convergence} confirms that the present immersed-boundary method exhibits a second-order convergence, with an exponent $p=1.912$.

\subsection{The turbulent flow in the human nose}
\label{sec:nose}

The air flow inside the human nose is an important application, owing to the obvious implications of a healthy breathing. 
The prevalence of anatomical malformations (like e.g. septal deviations, or hypertrophy of the turbinates) is huge, with nasal breathing difficulties affecting up to one third of the entire world population \cite{stewart-etal-2010}. 
The physiologically healthy flow through the nasal cavities is difficult to define, as no single flow feature can be shown to correlate with the perceived breathing quality.
In recent years the number of numerical studies dealing with the fluid mechanics of the human nose, built upon the patient-specific anatomic information provided by CT scans, has increased considerably. 
While the majority of such studies consists in simple RANS simulations executed with commercial, finite-volumes software, the availability of accurate reference solutions remains essential for validating physiology studies of fundamental character, and becomes clinically important whenever specific and unusual anatomies need to be evaluated for diagnosing and surgery planning. 
However, so far very few studies, e.g. Refs. \cite{calmet-etal-2016, li-etal-2017-nose}, have described the nose flow with DNS-like resolution, owing to the combination of its extreme geometrical complexity and the accompanying significant computational cost; none of them employed an immersed-boundary approach.

\begin{figure}
\centering
\includegraphics[width=0.7\textwidth]{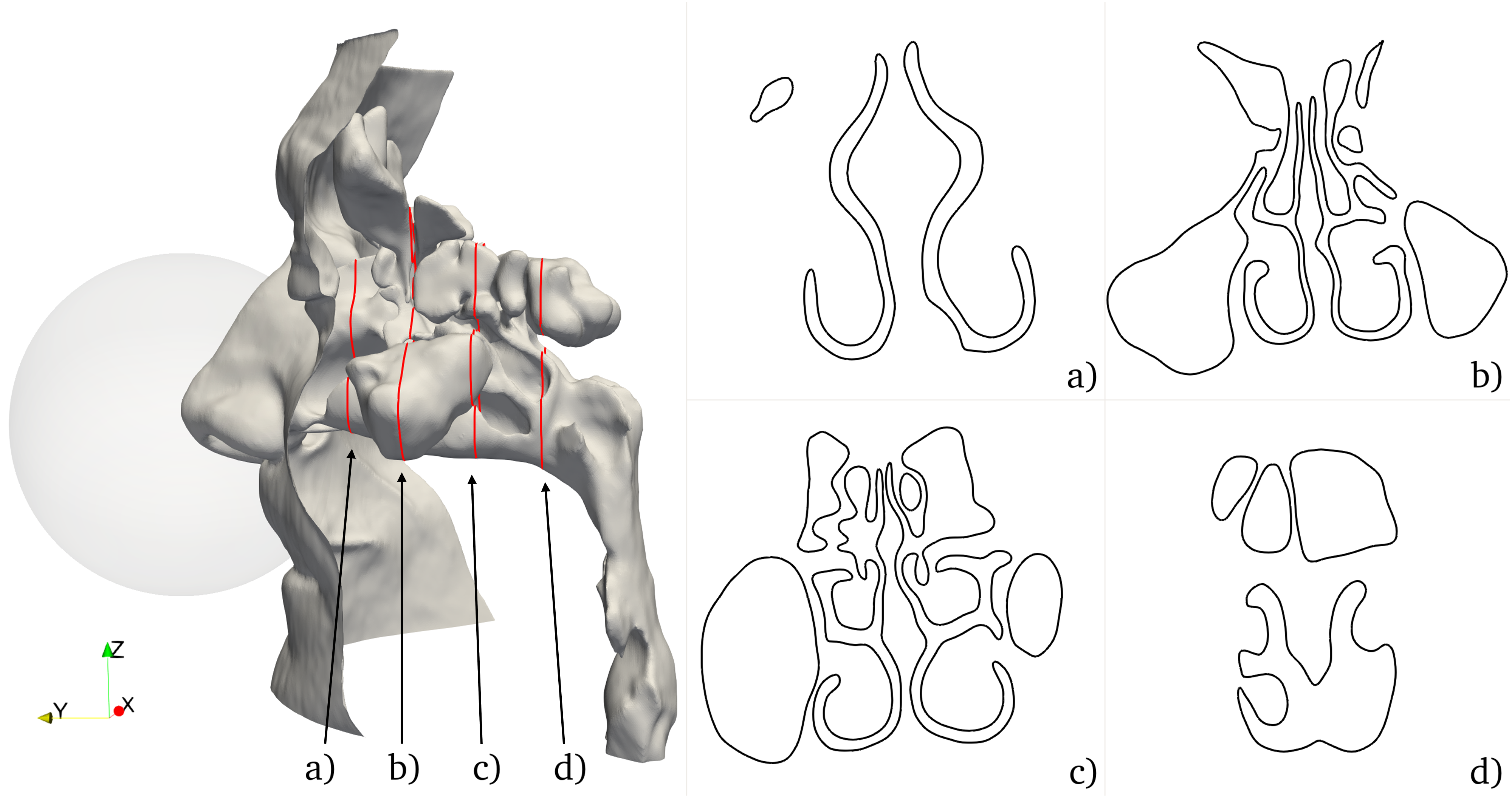}
\caption{Anatomy of the human airways, after a patient-specific reconstruction of a CT scan. The surface represents the boundary of the computational domain, and is augmented by a spherical volume placed outside the nose tip, whose goal is to set the computational boundary away from the nostrils. The vertical (coronal) sections on the right illustrate how the complex cross-sectional shape varies along the passageways, including the nasal vestibulum, the anterior part of the meati (a), the intermediate sections shaped by the turbinates (b) and (c), down to the rhinopharynx (d).}
\label{fig:anatomy}
\end{figure}

In the context of the present work, the nasal airflow represents an ideal test bed for the IBM, whose accuracy and computational performance can be assessed on a geometrically challenging scenario.
In this study, we consider therefore one specific sinonasal anatomy, that has been recently used in a tomo-PIV experiment described in Ref. \cite{tauwald-etal-2024}. The anatomy is derived  from segmentation of a CT scan of a healthy patient, composed of 384 DICOM images with sagittal and coronal resolution of $0.5 \ mm$ and an axial gap of $0.6 \ mm$.
The anatomy is segmented at constant radiodensity threshold, with the assistance of an experienced surgeon, according to a well established procedure \cite{quadrio-etal-2016}, to identify the interface between air and solid tissues. For experimental reasons, the whole anatomy has been then enlarged by a factor of 2. 
Figure \ref{fig:anatomy} portraits the anatomy of the internal nose, and demonstrates how the actual geometrical boundary is extremely complex, with evident three-dimensional features and the presence of large lateral volumes, the paranasal sinuses, which are only loosely connected to the main airways via small orifices called ostia. 
Figure \ref{fig:anatomy} also highlights, by means of coronal sections, how the cross-sectional shape of the airways varies significantly from the nose tip towards the throat.
The computational domain also comprises a spherical volume, shown in figure \ref{fig:anatomy}, which surrounds the external nose: the sphere is designed to locate the boundary of the computational domain far from the nostrils, while keeping the computational overhead within reason.

In this work a simple steady inspiration is considered, where a (constant in time) pressure drop is imposed between the inlet at the external surface of the sphere and the outlet at the trachea; the numerical value of $5 \, Pa$ and the corresponding volumetric flow rate of approximately $600 \, cm^3 / s$ correspond, after the factor-of-two geometrical expansion is accounted for with dynamic similarity arguments, to a mild physical activity \cite{wang-lee-gordon-2012}. Note that, because we are replicating an experimental study, dimensional quantities are used in this section.
Regardless of the time-independent boundary conditions, the flow is unsteady, with three-dimensional shear layers and vortical structures; in some regions the flow becomes turbulent.  
Starting from an initial condition of resting flow everywhere, the simulations are advanced in time until the initial transient has vanished, and then further integrated for about $1 \, s$ of physical time to compute time averages.

\begin{figure}
\centering
\includegraphics[width=0.49\textwidth]{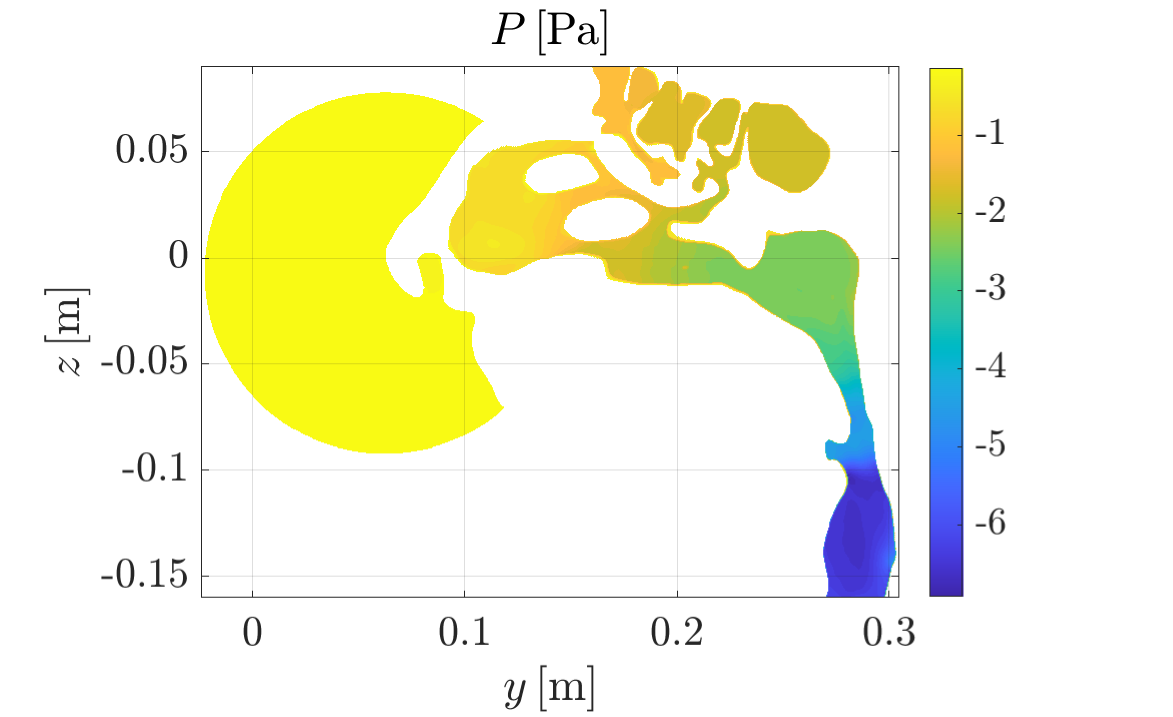}
\includegraphics[width=0.49\textwidth]{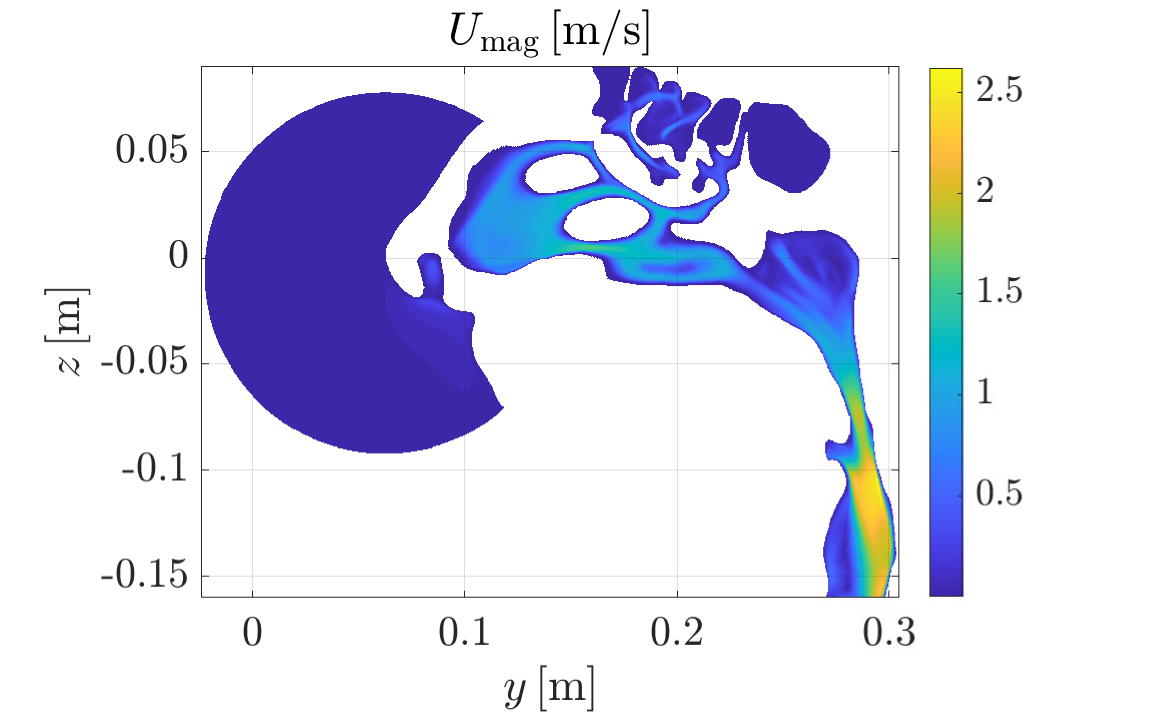}
\caption{Left: mean pressure field in a sagittal section. Right: mean velocity magnitude field in the same section.}
\label{fig:nose-mean}
\end{figure}

Some features of the time-averaged flow are illustrated in figure \ref{fig:nose-mean}. 
The left panel plots the mean pressure field in a representative sagittal section that cuts through the right passageway, and illustrates the pressure decrease from the outer ambient to the throat. The pressure drop is particularly localized and noticeable at the nasal valve and at the striction determined by the larynx.
The right panel plots the magnitude of the mean velocity vector in the same section. During inspiration the external air is first accelerated abruptly in correspondence of the nasal valve, and quite large velocity values are observed in certain portions of the meati; the airflow then enters the rhinopharynx with quite definite shear layers, and then transforms into a laryngeal jet after the narrowing at the larynx, where the largest velocities are found.

\begin{table}
\centering
\begin{tabular}{rrrrr}
$N_{tot}$  & $\Delta$ [$mm$]  & $Q$ [$cm^3/s$] & $\sigma$ [$cm^3/s$] \\
\hline
 $3749528$  & $0.9978$ & $598.591$ & $0.3306$ \\
 $6289066$  & $0.8317$ & $608.040$ & $0.3909$ \\ 
$10603652$  & $0.6934$ & $615.943$ & $0.3407$ \\
$17969591$  & $0.5775$ & $620.155$ & $0.2288$ \\
$30476246$  & $0.4815$ & $624.907$ & $0.5118$ \\ 
$51936820$  & $0.4011$ & $628.032$ & $0.6040$ \\ 
$91969807$  & $0.3343$ & $628.626$ & $0.9309$ \\ 
$157849674$ & $0.2787$ & $630.130$ & $0.6996$ \\ 
$273071473$ & $0.2322$ & $631.042$ & $0.5586$ \\ 
\end{tabular}
\caption{Spatial convergence study for the nose test case. The table lists the total number of grid points in the fluid volume, the (isotropic) grid spacing in millimeters, the value $Q$ of the temporally averaged flow rate at the trachea, and the root-mean-square value $\sigma$ of the variance of the estimate of the mean, computed after Ref. \cite{russo-luchini-2017}.}
\label{tab:nose-convergence}
\end{table}

\begin{figure}
\centering
\includegraphics[width=0.8\textwidth]{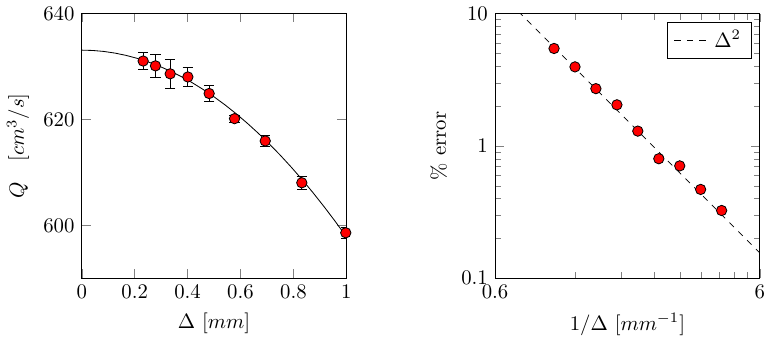}
\caption{Convergence of the flow rate for the nose test case. Left: variation of the time-average flow rate $Q$ with the (isotropic) spatial resolution $\Delta$; the error bars correspond to $\pm 3 \sigma$, where the root-mean-square value $\sigma$ of the uncertainty of the mean is computed after \cite{russo-luchini-2017}. The continuous line is a fit of equation \eqref{eq:conv}, yielding $Q^{(e)}=633.1 \, cm^3/s$, $C=3.502 \times 10^{-7}$ and $p=1.999$. Right: percentage error $100 (Q^{(e)}-Q)/Q^{(e)}$ of the mean flow rate versus spatial resolution. The dashed line shows the expected second-order decrease.}
\label{fig:nose-convergence}
\end{figure}

Table \ref{tab:nose-convergence} reports the results of a spatial convergence study, where the time-averaged value of the flow rate $Q$ at the trachea for the fixed pressure drop of $5 \, Pa$ is observed as the spatial resolution is changed. 
The calculations have been carried out for a fixed time step size $\Delta t = 0.02 \, ms$, that is small enough to ensure that the the time discretization error is not dominant. The geometrical complexity suggests the use of isotropic spacing, hence cubic cells are used with edge length $\Delta$. The coarsest mesh consists of about 3.7 millions points within the fluid volume; the size of its cells is $1 \, mm$ (which is comparable to the resolution of the CT scan, if the model scaling is considered) and still allows a decent representation of the smallest flow structures. The largest employed mesh has about 273 millions points and a spatial resolution of $230 \, \mu m$. 
Fitting the formula \eqref{eq:conv} to the data of table \ref{tab:nose-convergence}, following the same procedure discussed above in Sec. \ref{sec:wavy}, yields an exponent $p=1.999$: the second-order convergence is fully confirmed, as shown graphically by figure \ref{fig:nose-convergence}.

%% file: wavybottom-unstable.tex
\begin{tikzpicture}[scale=0.5]	

\draw[fill=gray!50, opacity=0.5] plot[smooth, samples=100, domain=4*pi:0] ({\x}, {0.5+0.5*cos(\x*180/pi)}) -- (0,-0.8) -- (4*pi,-0.8) -- (4*pi,1);
\draw[-] (0,1+4) -- (4*pi,1+4);
\draw[scale=1, domain=1:5, smooth, variable=\y, black] plot({-1.6875+1.8750*\y-0.1875*\y*\y},{\y});

\node at (0.5,3.5) {$U$};
\draw[->] (0,5)--(3,5);
\draw[->] (0,3)--(2.25,3);
\draw[->] (0,4)--(2.8125,4);
\draw[->] (0,2)--(1.3125,2);

\draw[dotted] (0,1)--(0,5);
\draw[dotted] (4*pi,1)--(4*pi,5);

\node at (2*pi,2.2) {\color{red}unstable};
\draw[<-] (2*pi-0.5,1.65)--(2*pi-2,1);
\draw[<-] (2*pi+0.5,1.65)--(2*pi+2,1);

\draw[->] (0, -5) -- (4*pi, -5);
\draw[->] (0,-5)  -- (0,-1);
\node at (2.8,-1.75) {wall shear stress};
\node at (10,-1.75) {\color{red}$\tau_{1s}<0$};
\draw[scale=1, domain=0:4*pi, smooth, variable=\x, black] plot ({\x}, {-3.5+0.5+0.5*cos(\x*180/pi+pi/2*180/pi)});
\end{tikzpicture}

%% file: wavybottom-stable.tex
\begin{tikzpicture}[scale=0.5]	
\node at (11.7,4.6) {$z=1$};
\draw[fill=gray!50, opacity=0.5] plot[smooth, samples=100, domain=4*pi:0] ({\x}, {0.5+0.5*cos(\x*180/pi)}) -- (0,-0.8) -- (4*pi,-0.8) -- (4*pi,1);
	\node at (9.5,-0.4) {$z_w = \epsilon(1+\cos (\alpha x) )$};
\draw[-] (0,1+4) -- (4*pi,1+4);
\draw[scale=1, domain=1:5, smooth, variable=\y, black] plot({-1.6875+1.8750*\y-0.1875*\y*\y},{\y});

\node at (0.5,3.5) {$U$};
\draw[->] (0,5)--(3,5);
\draw[->] (0,3)--(2.25,3);
\draw[->] (0,4)--(2.8125,4);
\draw[->] (0,2)--(1.3125,2);

\draw[dotted] (0,1)--(0,5);
\draw[dotted] (4*pi,1)--(4*pi,5);

\node at (2*pi,2.2) {\color{red}stable};
\draw[->] (2*pi-0.5,1.65)--(2*pi-2,1);
\draw[->] (2*pi+0.5,1.65)--(2*pi+2,1);

\draw[->] (0, -5) -- (4*pi, -5);
\draw[->] (0,-5)  -- (0,-1);
\node at (2.8,-1.75) {wall shear stress};
\node at (10,-1.75) {\color{red}$\tau_{1s}>0$};
\node at (2*pi,-4) {$\tau = \tau_0 + \tau_{1c} \cos(\alpha x) + \tau_{1s} \sin(\alpha x)$};
\draw[scale=1, domain=0:4*pi, smooth, variable=\x, black] plot ({\x}, {-3.5+0.5+0.5*cos(\x*180/pi-pi/2*180/pi)});
\end{tikzpicture}

%% file: conclusions.tex
\section{Conclusions}
\label{sec:conclusions}

We have presented a simple and efficient implicit second-order immersed boundary method (IBM) for the direct numerical simulation of the incompressible Navier--Stokes equations.

Our IBM belongs to the class of methods employing a discrete forcing; its computational efficiency descends from the tight integration with the underlying numerical discretization, based on second-order accurate central finite differences computed on a staggered Cartesian grid and on a generic explicit scheme for advancing the solution in time. 
Such integration is possible because the boundary-distance information is concentrated in a single point, namely the centre point of the stencil that discretizes the Laplacian operator. 
This allows the IBM correction to be made implicit at no cost: this is perhaps the main novelty of the present IBM implementation.

This type of correction offers crucial advantages. 
On the one hand, since the correction is applied only to the central point of the stencil, the formulation enjoys extreme simplicity in terms of implementation: the cobweb of IFs and {\em ad-hoc} coding instructions required to handle special cases when parts of the boundary come close to each other becomes unnecessary.
On the other hand, the computational cost of the IBM correction is brought to a minimum. In order to estimate it, a simple experiment consists in switching the correction off, thus reinstating the baseline first-order-accurate staircase approximation of the boundary; timing the execution of the solver with and without the correction quantifies the extra cost necessary to achieve second-order accuracy near the boundary. Such timing experiments led us to the observation that hardly any extra cost is visible at all.
The implicit nature of the IBM also has important favourable consequences on the stability of the numerical method. The explicit IBMs, in fact, fall into trouble whenever a grid point coincides or even approaches the boundary, whereas for the implicit IBM the solution monotonically tends to zero as it should.

The IBM has been introduced and described in conjunction with two simple linear problems, where the spatial and temporal accuracy of the method have been separately addressed. 
In the second part of the paper, the method has been first applied to a two-dimensional, steady and unsteady laminar cylinder flow. Then, two turbulent flows are considered, namely the channel flow over a wavy bottom and the air flow through the intricate anatomy of the human nose. A full second-order accuracy has been confirmed even in these examples, which contain the entire complexity of the Navier--Stokes equations. The IBM described in this paper lies at the heart of a DNS solver, currently under active development, which is aimed at both CPU and GPU architectures, and promises to achieve very interesting computational speed.

%% file: main.bbl
\begin{thebibliography}{52}
\expandafter\ifx\csname natexlab\endcsname\relax\def\natexlab#1{#1}\fi
\providecommand{\url}[1]{\texttt{#1}}
\providecommand{\href}[2]{#2}
\providecommand{\path}[1]{#1}
\providecommand{\DOIprefix}{doi:}
\providecommand{\ArXivprefix}{arXiv:}
\providecommand{\URLprefix}{URL: }
\providecommand{\Pubmedprefix}{pmid:}
\providecommand{\doi}[1]{\href{http://dx.doi.org/#1}{\path{#1}}}
\providecommand{\Pubmed}[1]{\href{pmid:#1}{\path{#1}}}
\providecommand{\bibinfo}[2]{#2}
\ifx\xfnm\relax \def\xfnm[#1]{\unskip,\space#1}\fi
\bibitem[{Balaras(2004)}]{balaras-2004}
\bibinfo{author}{Balaras, E.}, \bibinfo{year}{2004}.
\newblock \bibinfo{title}{Modeling complex boundaries using an external force
  field on fixed {{Cartesian}} grids in large-eddy simulations}.
\newblock \bibinfo{journal}{Computers \& Fluids} \bibinfo{volume}{33},
  \bibinfo{pages}{375--404}.
\newblock \DOIprefix\doi{10.1016/S0045-7930(03)00058-6}.
\bibitem[{Blondeaux(1990)}]{blondeaux-1990}
\bibinfo{author}{Blondeaux, P.}, \bibinfo{year}{1990}.
\newblock \bibinfo{title}{Sand ripples under sea waves {{Part}} 1. {{Ripple}}
  formation}.
\newblock \bibinfo{journal}{J. Fluid Mech.} \bibinfo{volume}{218},
  \bibinfo{pages}{1--17}.
\newblock \DOIprefix\doi{10.1017/S0022112090000908}.
\bibitem[{Calmet et~al.(2016)Calmet, Gambaruto, Bates, V{\'a}zquez, Houzeaux
  and Doorly}]{calmet-etal-2016}
\bibinfo{author}{Calmet, H.}, \bibinfo{author}{Gambaruto, A.},
  \bibinfo{author}{Bates, A.}, \bibinfo{author}{V{\'a}zquez, M.},
  \bibinfo{author}{Houzeaux, G.}, \bibinfo{author}{Doorly, D.},
  \bibinfo{year}{2016}.
\newblock \bibinfo{title}{Large-scale {{CFD}} simulations of the transitional
  and turbulent regime for the large human airways during rapid inhalation}.
\newblock \bibinfo{journal}{Comp. Biol. Med.} \bibinfo{volume}{69},
  \bibinfo{pages}{166--180}.
\newblock \DOIprefix\doi{10.1016/j.compbiomed.2015.12.003}.
\bibitem[{Charru et~al.(2013)Charru, Andreotti and Claudin}]{charru-etal-2013}
\bibinfo{author}{Charru, F.}, \bibinfo{author}{Andreotti, B.},
  \bibinfo{author}{Claudin, P.}, \bibinfo{year}{2013}.
\newblock \bibinfo{title}{Sand ripples and dunes}.
\newblock \bibinfo{journal}{Annu. Rev. Fluid Mech.} \bibinfo{volume}{45},
  \bibinfo{pages}{469--493}.
\newblock \DOIprefix\doi{10.1146/annurev-fluid-011212-140806},
  \href{http://arxiv.org/abs/https://doi.org/10.1146/annurev-fluid-011212-140806}{\tt
  arXiv:https://doi.org/10.1146/annurev-fluid-011212-140806}.
\bibitem[{Chi et~al.(2020)Chi, Abdelsamie and Th{\'e}venin}]{chi-etal-2020}
\bibinfo{author}{Chi, C.}, \bibinfo{author}{Abdelsamie, A.},
  \bibinfo{author}{Th{\'e}venin, D.}, \bibinfo{year}{2020}.
\newblock \bibinfo{title}{A directional ghost-cell immersed boundary method for
  incompressible flows}.
\newblock \bibinfo{journal}{J. Comp. Phys.} \bibinfo{volume}{404},
  \bibinfo{pages}{109122}.
\newblock \DOIprefix\doi{10.1016/j.jcp.2019.109122}.
\bibitem[{Chi et~al.(2017)Chi, Lee and Im}]{chi-lee-im-2017}
\bibinfo{author}{Chi, C.}, \bibinfo{author}{Lee, B.J.}, \bibinfo{author}{Im,
  H.G.}, \bibinfo{year}{2017}.
\newblock \bibinfo{title}{An improved ghost-cell immersed boundary method for
  compressible flow simulations}.
\newblock \bibinfo{journal}{Int. J. Num. Meth. Fluids} \bibinfo{volume}{83},
  \bibinfo{pages}{132--148}.
\newblock \DOIprefix\doi{10.1002/fld.4262}.
\bibitem[{Coutanceau and Bouard(1977)}]{coutanceau-bouard-1977}
\bibinfo{author}{Coutanceau, M.}, \bibinfo{author}{Bouard, R.},
  \bibinfo{year}{1977}.
\newblock \bibinfo{title}{Experimental determination of the main features of
  the viscous flow in the wake of a circular cylinder in uniform translation.
  {{Part}} 1. {{Steady}} flow}.
\newblock \bibinfo{journal}{J. Fluid Mech.} \bibinfo{volume}{79},
  \bibinfo{pages}{231--256}.
\newblock \DOIprefix\doi{10.1017/S0022112077000135}.
\bibitem[{{de Tullio} and Pascazio(2016)}]{detullio-pascazio-2016}
\bibinfo{author}{{de Tullio}, M.D.}, \bibinfo{author}{Pascazio, G.},
  \bibinfo{year}{2016}.
\newblock \bibinfo{title}{A moving-least-squares immersed boundary method for
  simulating the fluid--structure interaction of elastic bodies with arbitrary
  thickness}.
\newblock \bibinfo{journal}{J. Comp. Phys.} \bibinfo{volume}{325},
  \bibinfo{pages}{201--225}.
\newblock \DOIprefix\doi{10.1016/j.jcp.2016.08.020}.
\bibitem[{Fadlun et~al.(2000)Fadlun, Verzicco, Orlandi and
  {Mohd-Yusof}}]{fadlun-etal-2000}
\bibinfo{author}{Fadlun, E.A.}, \bibinfo{author}{Verzicco, R.},
  \bibinfo{author}{Orlandi, P.}, \bibinfo{author}{{Mohd-Yusof}, J.},
  \bibinfo{year}{2000}.
\newblock \bibinfo{title}{Combined {{Immersed-Boundary Finite-Difference
  Methods}} for {{Three-Dimensional Complex Flow Simulations}}}.
\newblock \bibinfo{journal}{J. Comp. Phys.} \bibinfo{volume}{161},
  \bibinfo{pages}{35--60}.
\newblock \DOIprefix\doi{10.1006/jcph.2000.6484}.
\bibitem[{Fauci and Peskin(1988)}]{fauci-peskin-1988}
\bibinfo{author}{Fauci, L.J.}, \bibinfo{author}{Peskin, C.S.},
  \bibinfo{year}{1988}.
\newblock \bibinfo{title}{A computational model of aquatic animal locomotion}.
\newblock \bibinfo{journal}{J. Comp. Phys.} \bibinfo{volume}{77},
  \bibinfo{pages}{85--108}.
\newblock \DOIprefix\doi{10.1016/0021-9991(88)90158-1}.
\bibitem[{Fedkiw(2002)}]{fedkiw-2002}
\bibinfo{author}{Fedkiw, R.}, \bibinfo{year}{2002}.
\newblock \bibinfo{title}{Coupling an {{Eulerian Fluid Calculation}} to a
  {{Lagrangian Solid Calculation}} with the {{Ghost Fluid Method}}}.
\newblock \bibinfo{journal}{J. Comp. Phys.} \bibinfo{volume}{175},
  \bibinfo{pages}{200--224}.
\newblock \DOIprefix\doi{10.1006/jcph.2001.6935}.
\bibitem[{Gao et~al.(2007)Gao, Tseng and Lu}]{gao-tseng-lu-2007}
\bibinfo{author}{Gao, T.}, \bibinfo{author}{Tseng, Y.H.}, \bibinfo{author}{Lu,
  X.Y.}, \bibinfo{year}{2007}.
\newblock \bibinfo{title}{An improved hybrid {{Cartesian}}/immersed boundary
  method for fluid--solid flows}.
\newblock \bibinfo{journal}{Int. J. Num. Meth. Fluids} \bibinfo{volume}{55},
  \bibinfo{pages}{1189--1211}.
\newblock \DOIprefix\doi{10.1002/fld.1522}.
\bibitem[{Ghias et~al.(2007)Ghias, Mittal and Dong}]{ghias-mittal-dong-2007}
\bibinfo{author}{Ghias, R.}, \bibinfo{author}{Mittal, R.},
  \bibinfo{author}{Dong, H.}, \bibinfo{year}{2007}.
\newblock \bibinfo{title}{A sharp interface immersed boundary method for
  compressible viscous flows}.
\newblock \bibinfo{journal}{J. Comp. Phys.} \bibinfo{volume}{225},
  \bibinfo{pages}{528--553}.
\newblock \DOIprefix\doi{10.1016/j.jcp.2006.12.007}.
\bibitem[{Gibou et~al.(2002)Gibou, Fedkiw, Cheng and Kang}]{gibou-etal-2002}
\bibinfo{author}{Gibou, F.}, \bibinfo{author}{Fedkiw, R.},
  \bibinfo{author}{Cheng, L.T.}, \bibinfo{author}{Kang, M.},
  \bibinfo{year}{2002}.
\newblock \bibinfo{title}{A {{Second-Order-Accurate Symmetric Discretization}}
  of the {{Poisson Equation}} on {{Irregular Domains}}}.
\newblock \bibinfo{journal}{J. Comp. Phys.} \bibinfo{volume}{176},
  \bibinfo{pages}{205--227}.
\newblock \DOIprefix\doi{10.1006/jcph.2001.6977}.
\bibitem[{Goldstein et~al.(1993)Goldstein, Handler and
  Sirovich}]{goldstein-handler-sirovich-1993}
\bibinfo{author}{Goldstein, D.}, \bibinfo{author}{Handler, R.},
  \bibinfo{author}{Sirovich, L.}, \bibinfo{year}{1993}.
\newblock \bibinfo{title}{Modeling a {{No-Slip Flow Boundary}} with an
  {{External Force Field}}}.
\newblock \bibinfo{journal}{J. Comput. Phys.} \bibinfo{volume}{105},
  \bibinfo{pages}{354--366}.
\newblock \DOIprefix\doi{10.1006/jcph.1993.1081}.
\bibitem[{Griffith and Patankar(2020)}]{griffith-patankar-2020}
\bibinfo{author}{Griffith, B.}, \bibinfo{author}{Patankar, N.A.},
  \bibinfo{year}{2020}.
\newblock \bibinfo{title}{Immersed {{Methods}} for {{Fluid}}--{{Structure
  Interaction}}}.
\newblock \bibinfo{journal}{Annu. Rev. Fluid Mech.} \bibinfo{volume}{52},
  \bibinfo{pages}{421--448}.
\newblock \DOIprefix\doi{10.1146/annurev-fluid-010719-060228}.
\bibitem[{Guilmineau and Queutey(2002)}]{guilmineau-queteuy-2002}
\bibinfo{author}{Guilmineau, E.}, \bibinfo{author}{Queutey, P.},
  \bibinfo{year}{2002}.
\newblock \bibinfo{title}{A numerical simulation of vortex shedding from an
  oscillating cyrcular cylinder}.
\newblock \bibinfo{journal}{J. Fluids Struct.} \bibinfo{volume}{16},
  \bibinfo{pages}{773--794}.
\newblock \DOIprefix\doi{10.1006/jfls.2002.0449}.
\bibitem[{Hecht(2012)}]{hecht-2012}
\bibinfo{author}{Hecht, F.}, \bibinfo{year}{2012}.
\newblock \bibinfo{title}{New development in {{FreeFem}}++}.
\newblock \bibinfo{journal}{J. Numer. Math.} \bibinfo{volume}{20}.
\newblock \DOIprefix\doi{10.1515/jnum-2012-0013}.
\bibitem[{Iaccarino and Verzicco(2003)}]{iaccarino-verzicco-2003}
\bibinfo{author}{Iaccarino, G.}, \bibinfo{author}{Verzicco, R.},
  \bibinfo{year}{2003}.
\newblock \bibinfo{title}{Immersed boundary technique for turbulent flow
  simulations}.
\newblock \bibinfo{journal}{Appl. Mech. Rev.} \bibinfo{volume}{56},
  \bibinfo{pages}{331--347}.
\newblock \DOIprefix\doi{10.1115/1.1563627}.
\bibitem[{Kim et~al.(1987)Kim, Moin and Moser}]{kim-moin-moser-1987}
\bibinfo{author}{Kim, J.}, \bibinfo{author}{Moin, P.}, \bibinfo{author}{Moser,
  R.}, \bibinfo{year}{1987}.
\newblock \bibinfo{title}{Turbulence statistics in fully developed channel flow
  at low {{Reynolds}} number}.
\newblock \bibinfo{journal}{J. Fluid Mech.} \bibinfo{volume}{177},
  \bibinfo{pages}{133--166}.
\newblock \DOIprefix\doi{10.1017/S0022112087000892}.
\bibitem[{Kim and Choi(2019)}]{kim-choi-2019}
\bibinfo{author}{Kim, W.}, \bibinfo{author}{Choi, H.}, \bibinfo{year}{2019}.
\newblock \bibinfo{title}{Immersed boundary methods for fluid-structure
  interaction: {{A}} review}.
\newblock \bibinfo{journal}{International Journal of Heat and Fluid Flow}
  \bibinfo{volume}{75}, \bibinfo{pages}{301--309}.
\newblock \DOIprefix\doi{10.1016/j.ijheatfluidflow.2019.01.010}.
\bibitem[{Kim and Lai(2010)}]{kim-lai-2010}
\bibinfo{author}{Kim, Y.}, \bibinfo{author}{Lai, M.C.}, \bibinfo{year}{2010}.
\newblock \bibinfo{title}{Simulating the dynamics of inextensible vesicles by
  the penalty immersed boundary method}.
\newblock \bibinfo{journal}{J. Comput. Phys.} \bibinfo{volume}{229},
  \bibinfo{pages}{4840--4853}.
\bibitem[{Kim and Peskin(2007)}]{kim-peskin-2007}
\bibinfo{author}{Kim, Y.}, \bibinfo{author}{Peskin, C.}, \bibinfo{year}{2007}.
\newblock \bibinfo{title}{Penalty immersed boundary method for an elastic
  boundary with mass}.
\newblock \bibinfo{journal}{Phys. Fluids} \bibinfo{volume}{19},
  \bibinfo{pages}{053103}.
\newblock \DOIprefix\doi{10.1063/1.2734674}.
\bibitem[{Li et~al.(2017)Li, Jiang, Dong and Zhao}]{li-etal-2017-nose}
\bibinfo{author}{Li, C.}, \bibinfo{author}{Jiang, J.}, \bibinfo{author}{Dong,
  H.}, \bibinfo{author}{Zhao, K.}, \bibinfo{year}{2017}.
\newblock \bibinfo{title}{Computational modeling and validation of human nasal
  airflow under various breathing conditions}.
\newblock \bibinfo{journal}{J. Biomech.} \bibinfo{volume}{64},
  \bibinfo{pages}{59--68}.
\newblock \DOIprefix\doi{10.1016/j.jbiomech.2017.08.031}.
\bibitem[{Li et~al.(2023)Li, Bale, Wang and Tsubokura}]{li-etal-2023}
\bibinfo{author}{Li, C.G.}, \bibinfo{author}{Bale, R.}, \bibinfo{author}{Wang,
  W.}, \bibinfo{author}{Tsubokura, M.}, \bibinfo{year}{2023}.
\newblock \bibinfo{title}{A sharp interface immersed boundary method for
  thin-walled geometries in viscous compressible flows}.
\newblock \bibinfo{journal}{Int. J. Mech. Sci.} \bibinfo{volume}{253},
  \bibinfo{pages}{108401}.
\newblock \DOIprefix\doi{10.1016/j.ijmecsci.2023.108401}.
\bibitem[{Lu and Dalton(1996)}]{lu-dalton-1996}
\bibinfo{author}{Lu, X.Y.}, \bibinfo{author}{Dalton, C.}, \bibinfo{year}{1996}.
\newblock \bibinfo{title}{Calculation of the timing of vortex formation from an
  oscillating cylinder}.
\newblock \bibinfo{journal}{J. Fluids Struct.} \bibinfo{volume}{10},
  \bibinfo{pages}{527--541}.
\newblock \DOIprefix\doi{10.1006/jfls.1996.0035}.
\bibitem[{Luchini(2016)}]{luchini-2016}
\bibinfo{author}{Luchini, P.}, \bibinfo{year}{2016}.
\newblock \bibinfo{title}{Immersed-boundary simulations of turbulent flow past
  a sinusoidally undulated river bottom}.
\newblock \bibinfo{journal}{Eur. J. Mech. B/Fluids} \bibinfo{volume}{55},
  \bibinfo{pages}{340--347}.
\newblock \DOIprefix\doi{10.1016/j.euromechflu.2015.08.007}.
\bibitem[{Mittal et~al.(2008)Mittal, Dong, Bozkurttas, Najjar, Vargas and {von
  Loebbecke}}]{mittal-etal-2008}
\bibinfo{author}{Mittal, R.}, \bibinfo{author}{Dong, H.},
  \bibinfo{author}{Bozkurttas, M.}, \bibinfo{author}{Najjar, F.},
  \bibinfo{author}{Vargas, A.}, \bibinfo{author}{{von Loebbecke}, A.},
  \bibinfo{year}{2008}.
\newblock \bibinfo{title}{A versatile sharp interface immersed boundary method
  for incompressible flows with complex boundaries}.
\newblock \bibinfo{journal}{J. Comp. Phys.} \bibinfo{volume}{227},
  \bibinfo{pages}{4825--4852}.
\newblock \DOIprefix\doi{10.1016/j.jcp.2008.01.028}.
\bibitem[{Mittal and Iaccarino(2005)}]{mittal-iaccarino-2005}
\bibinfo{author}{Mittal, R.}, \bibinfo{author}{Iaccarino, G.},
  \bibinfo{year}{2005}.
\newblock \bibinfo{title}{Immersed {{Boundary Methods}}}.
\newblock \bibinfo{journal}{Annu. Rev. Fluid Mech.} \bibinfo{volume}{37},
  \bibinfo{pages}{239--261}.
\newblock \DOIprefix\doi{10.1146/annurev.fluid.37.061903.175743}.
\bibitem[{Mittal and Seo(2023)}]{mittal-seo-2023}
\bibinfo{author}{Mittal, R.}, \bibinfo{author}{Seo, J.}, \bibinfo{year}{2023}.
\newblock \bibinfo{title}{Origin and evolution of immersed boundary methods in
  computational fluid dynamics}.
\newblock \bibinfo{journal}{Phys. Rev. Fluids} \bibinfo{volume}{8},
  \bibinfo{pages}{100501}.
\newblock \DOIprefix\doi{10.1103/PhysRevFluids.8.100501}.
\bibitem[{Orlandi and Leonardi(2006)}]{orlandi-leonardi-2006}
\bibinfo{author}{Orlandi, P.}, \bibinfo{author}{Leonardi, S.},
  \bibinfo{year}{2006}.
\newblock \bibinfo{title}{{{DNS}} of turbulent channel flows with two- and
  three-dimensional roughness}.
\newblock \bibinfo{journal}{J. Turbul.} \bibinfo{volume}{7},
  \bibinfo{pages}{N73}.
\newblock \DOIprefix\doi{10.1080/14685240600827526}.
\bibitem[{Pan and Shen(2009)}]{pan-shen-2009}
\bibinfo{author}{Pan, D.}, \bibinfo{author}{Shen, T.T.}, \bibinfo{year}{2009}.
\newblock \bibinfo{title}{Computation of incompressible flows with immersed
  bodies by a simple ghost cell method}.
\newblock \bibinfo{journal}{Int. J. Num. Meth. Fluids} \bibinfo{volume}{60},
  \bibinfo{pages}{1378--1401}.
\newblock \DOIprefix\doi{10.1002/fld.1942}.
\bibitem[{Peskin(1972)}]{peskin-1972}
\bibinfo{author}{Peskin, C.}, \bibinfo{year}{1972}.
\newblock \bibinfo{title}{Flow patterns around heart valves: {{A}} numerical
  method}.
\newblock \bibinfo{journal}{J. Comp. Phys.} \bibinfo{volume}{10},
  \bibinfo{pages}{252--271}.
\newblock \DOIprefix\doi{10.1016/0021-9991(72)90065-4}.
\bibitem[{Pinelli et~al.(2010)Pinelli, Naqavi, Piomelli and
  Favier}]{pinelli-etal-2010}
\bibinfo{author}{Pinelli, A.}, \bibinfo{author}{Naqavi, I.Z.},
  \bibinfo{author}{Piomelli, U.}, \bibinfo{author}{Favier, J.},
  \bibinfo{year}{2010}.
\newblock \bibinfo{title}{Immersed-boundary methods for general
  finite-difference and finite-volume {{Navier}}--{{Stokes}} solvers}.
\newblock \bibinfo{journal}{J. Comp. Phys.} \bibinfo{volume}{229},
  \bibinfo{pages}{9073--9091}.
\newblock \DOIprefix\doi{10.1016/j.jcp.2010.08.021}.
\bibitem[{Quadrio et~al.(2016)Quadrio, Pipolo, Corti, Messina, Pesci, Saibene,
  Zampini and Felisati}]{quadrio-etal-2016}
\bibinfo{author}{Quadrio, M.}, \bibinfo{author}{Pipolo, C.},
  \bibinfo{author}{Corti, S.}, \bibinfo{author}{Messina, F.},
  \bibinfo{author}{Pesci, C.}, \bibinfo{author}{Saibene, A.},
  \bibinfo{author}{Zampini, S.}, \bibinfo{author}{Felisati, G.},
  \bibinfo{year}{2016}.
\newblock \bibinfo{title}{Effect of {{CT}} resolution and radiodensity
  threshold on the {{CFD}} evaluation of nasal airflow}.
\newblock \bibinfo{journal}{Med Biol Eng Comput} \bibinfo{volume}{54},
  \bibinfo{pages}{411--419}.
\bibitem[{Rai and Moin(1991)}]{rai-moin-1991}
\bibinfo{author}{Rai, M.}, \bibinfo{author}{Moin, P.}, \bibinfo{year}{1991}.
\newblock \bibinfo{title}{Direct simulations of turbulent flow using
  finite-difference schemes}.
\newblock \bibinfo{journal}{J. Comp. Phys.} \bibinfo{volume}{96},
  \bibinfo{pages}{15}.
\bibitem[{Russo and Luchini(2017)}]{russo-luchini-2017}
\bibinfo{author}{Russo, S.}, \bibinfo{author}{Luchini, P.},
  \bibinfo{year}{2017}.
\newblock \bibinfo{title}{A fast algorithm for the estimation of statistical
  error in {{DNS}} (or experimental) time averages}.
\newblock \bibinfo{journal}{J. Comput. Phys.} \bibinfo{volume}{347},
  \bibinfo{pages}{328--340}.
\bibitem[{Saiki and Biringen(1996)}]{saiki-biringen-1996}
\bibinfo{author}{Saiki, E.M.}, \bibinfo{author}{Biringen, S.},
  \bibinfo{year}{1996}.
\newblock \bibinfo{title}{Numerical {{Simulation}} of a {{Cylinder}} in
  {{Uniform Flow}}: {{Application}} of a {{Virtual Boundary Method}}}.
\newblock \bibinfo{journal}{J. Comp. Phys.} \bibinfo{volume}{123},
  \bibinfo{pages}{450--465}.
\newblock \DOIprefix\doi{10.1006/jcph.1996.0036}.
\bibitem[{Seo and Mittal(2011)}]{seo-mittal-2011}
\bibinfo{author}{Seo, J.}, \bibinfo{author}{Mittal, R.}, \bibinfo{year}{2011}.
\newblock \bibinfo{title}{A sharp-interface immersed boundary method with
  improved mass conservation and reduced spurious pressure oscillations}.
\newblock \bibinfo{journal}{J. Comp. Phys.} \bibinfo{volume}{230},
  \bibinfo{pages}{7347--7363}.
\newblock \DOIprefix\doi{10.1016/j.jcp.2011.06.003}.
\bibitem[{Sotiropoulos and Yang(2014)}]{sotiropoulos-yang-2014}
\bibinfo{author}{Sotiropoulos, F.}, \bibinfo{author}{Yang, X.},
  \bibinfo{year}{2014}.
\newblock \bibinfo{title}{Immersed boundary methods for simulating
  fluid--structure interaction}.
\newblock \bibinfo{journal}{Prog. Aero. Sci.} \bibinfo{volume}{65},
  \bibinfo{pages}{1--21}.
\newblock \DOIprefix\doi{10.1016/j.paerosci.2013.09.003}.
\bibitem[{Stewart et~al.(2010)Stewart, Thompson, Leweke and
  Hourigan}]{stewart-etal-2010}
\bibinfo{author}{Stewart, B.E.}, \bibinfo{author}{Thompson, M.C.},
  \bibinfo{author}{Leweke, T.}, \bibinfo{author}{Hourigan, K.},
  \bibinfo{year}{2010}.
\newblock \bibinfo{title}{The wake behind a cylinder rolling on a wall at
  varying rotation rates}.
\newblock \bibinfo{journal}{J. Fluid Mech.} \bibinfo{volume}{648},
  \bibinfo{pages}{225--256}.
\newblock \DOIprefix\doi{10.1017/S0022112009993053}.
\bibitem[{Tauwald et~al.(2024)Tauwald, Erzinger, Quadrio, R{\"u}tten, Stemmer
  and Krenkel}]{tauwald-etal-2024}
\bibinfo{author}{Tauwald, S.}, \bibinfo{author}{Erzinger, F.},
  \bibinfo{author}{Quadrio, M.}, \bibinfo{author}{R{\"u}tten, M.},
  \bibinfo{author}{Stemmer, C.}, \bibinfo{author}{Krenkel, L.},
  \bibinfo{year}{2024}.
\newblock \bibinfo{title}{Tomo-{{PIV}} in a patient-specific model of human
  nasal cavities: A methodological approach}.
\newblock \bibinfo{journal}{Meas. Sci. Technol.} \bibinfo{volume}{35},
  \bibinfo{pages}{055203}.
\newblock \DOIprefix\doi{10.1088/1361-6501/ad282c}.
\bibitem[{Tritton(1959)}]{tritton-1959}
\bibinfo{author}{Tritton, D.J.}, \bibinfo{year}{1959}.
\newblock \bibinfo{title}{Experiments on the flow past a circular cylinder at
  low {{Reynolds}} numbers}.
\newblock \bibinfo{journal}{J. Fluid Mech.} \bibinfo{volume}{6},
  \bibinfo{pages}{547--567}.
\newblock \DOIprefix\doi{10.1017/S0022112059000829}.
\bibitem[{Tseng and Ferziger(2003)}]{tseng-ferziger-2003}
\bibinfo{author}{Tseng, Y.H.}, \bibinfo{author}{Ferziger, J.},
  \bibinfo{year}{2003}.
\newblock \bibinfo{title}{A ghost-cell immersed boundary method for flow in
  complex geometry}.
\newblock \bibinfo{journal}{J. Comp. Phys.} \bibinfo{volume}{192},
  \bibinfo{pages}{593--623}.
\newblock \DOIprefix\doi{10.1016/j.jcp.2003.07.024}.
\bibitem[{Uhlmann(2005)}]{uhlmann-2005}
\bibinfo{author}{Uhlmann, M.}, \bibinfo{year}{2005}.
\newblock \bibinfo{title}{An immersed boundary method with direct forcing for
  the simulation of particulate flows}.
\newblock \bibinfo{journal}{J. Comp. Phys.} \bibinfo{volume}{209},
  \bibinfo{pages}{448--476}.
\newblock \DOIprefix\doi{10.1016/j.jcp.2005.03.017}.
\bibitem[{Vanella and Balaras(2009)}]{vanella-balaras-2009}
\bibinfo{author}{Vanella, M.}, \bibinfo{author}{Balaras, E.},
  \bibinfo{year}{2009}.
\newblock \bibinfo{title}{A moving-least-squares reconstruction for
  embedded-boundary formulations}.
\newblock \bibinfo{journal}{J. Comp. Phys.} \bibinfo{volume}{228},
  \bibinfo{pages}{6617--6628}.
\newblock \DOIprefix\doi{10.1016/j.jcp.2009.06.003}.
\bibitem[{Verzicco(2023)}]{verzicco-2023}
\bibinfo{author}{Verzicco, R.}, \bibinfo{year}{2023}.
\newblock \bibinfo{title}{Immersed {{Boundary Methods}}: {{Historical
  Perspective}} and {{Future Outlook}}}.
\newblock \bibinfo{journal}{Annu. Rev. Fluid Mech.} \bibinfo{volume}{55},
  \bibinfo{pages}{129--155}.
\newblock \DOIprefix\doi{10.1146/annurev-fluid-120720-022129}.
\bibitem[{Wang et~al.(2012)Wang, Lee and Gordon}]{wang-lee-gordon-2012}
\bibinfo{author}{Wang, D.}, \bibinfo{author}{Lee, H.}, \bibinfo{author}{Gordon,
  R.}, \bibinfo{year}{2012}.
\newblock \bibinfo{title}{Impacts of {{Fluid Dynamics Simulation}} in {{Study}}
  of {{Nasal Airflow Physiology}} and {{Pathophysiology}} in {{Realistic Human
  Three-Dimensional Nose Models}}}.
\newblock \bibinfo{journal}{Clin. Exper. Otorhinolaryngology}
  \bibinfo{volume}{5}, \bibinfo{pages}{181--187}.
\bibitem[{Williamson(1988)}]{williamson-1988}
\bibinfo{author}{Williamson, C.H.K.}, \bibinfo{year}{1988}.
\newblock \bibinfo{title}{Defining a universal and continuous
  {{Strouhal}}--{{Reynolds}} number relationship for the laminar vortex
  shedding of a circular cylinder}.
\newblock \bibinfo{journal}{Phys. Fluids} \bibinfo{volume}{31},
  \bibinfo{pages}{2742--2744}.
\newblock \DOIprefix\doi{10.1063/1.866978}.
\bibitem[{Ye et~al.(1999)Ye, Mittal, Udaykumar and Shyy}]{ye-etal-1999}
\bibinfo{author}{Ye, T.}, \bibinfo{author}{Mittal, R.},
  \bibinfo{author}{Udaykumar, H.S.}, \bibinfo{author}{Shyy, W.},
  \bibinfo{year}{1999}.
\newblock \bibinfo{title}{An {{Accurate Cartesian Grid Method}} for {{Viscous
  Incompressible Flows}} with {{Complex Immersed Boundaries}}}.
\newblock \bibinfo{journal}{J. Comp. Phys.} \bibinfo{volume}{156},
  \bibinfo{pages}{209--240}.
\newblock \DOIprefix\doi{10.1006/jcph.1999.6356}.
\bibitem[{Zhu and Peskin(2002)}]{zhu-peskin-2002}
\bibinfo{author}{Zhu, L.}, \bibinfo{author}{Peskin, C.}, \bibinfo{year}{2002}.
\newblock \bibinfo{title}{Simulation of a {{Flapping Flexible Filament}} in a
  {{Flowing Soap Film}} by the {{Immersed Boundary Method}}}.
\newblock \bibinfo{journal}{J. Comp. Phys.} \bibinfo{volume}{179},
  \bibinfo{pages}{452--468}.
\newblock \DOIprefix\doi{10.1006/jcph.2002.7066}.
\bibitem[{Zhu et~al.(2024)Zhu, Chen, Chong, Lohse and Verzicco}]{zhu-etal-2024}
\bibinfo{author}{Zhu, X.}, \bibinfo{author}{Chen, Y.}, \bibinfo{author}{Chong,
  K.}, \bibinfo{author}{Lohse, D.}, \bibinfo{author}{Verzicco, R.},
  \bibinfo{year}{2024}.
\newblock \bibinfo{title}{A boundary condition-enhanced direct-forcing immersed
  boundary method for simulations of three-dimensional phoretic particles in
  incompressible flows}.
\newblock \bibinfo{journal}{J. Comp. Phys.} \bibinfo{volume}{509},
  \bibinfo{pages}{113028}.
\newblock \DOIprefix\doi{10.1016/j.jcp.2024.113028}.

\end{thebibliography}
